\documentclass[12pt,preprint]{aastex}

\def\lesssim{\mathrel{\hbox{\rlap{\hbox{\lower4pt\hbox{$\sim$}}}\hbox{$<$}}}}
\def\gtrsim{\mathrel{\hbox{\rlap{\hbox{\lower4pt\hbox{$\sim$}}}\hbox{$>$}}}}

\newcommand{\mbfB}{\mathbf{B}}
\newcommand{\mbfv}{\mathbf{v}}
\newcommand{\mbfJ}{\mathbf{J}}
\newcommand{\mbfk}{\mathbf{k}}
\newcommand{\f}   {\frac}
\newcommand{\del}{\nabla}

\begin{document}

\title{Evolution of Unmagnetized and Magnetized Shear Layers}

\shorttitle{Evolution of Shear Layers}

\author{M. L. Palotti\altaffilmark{1,}\altaffilmark{4}}
\author{F. Heitsch\altaffilmark{3}}
\author{E. G. Zweibel\altaffilmark{1,}\altaffilmark{2,}\altaffilmark{4}}
\author{Y.-M. Huang\altaffilmark{1,}\altaffilmark{4}}
\altaffiltext{1}{Department of Physics, U Wisconsin-Madison 1150 University Ave. Madison, WI 53706}
\altaffiltext{2}{Department of Astronomy, U Wisconsin-Madison 475 North Charter St. Madison, WI 53706}
\altaffiltext{3}{Department of Astronomy, U Michigan-Ann Arbor 500 Church St. Ann Arbor, MI 48109}
\altaffiltext{4}{Center for Magnetic Self Organization in Laboratory and Astrophysical Plasmas}

\shortauthors{Palotti et al.}

\begin{abstract}
We present numerical simulations of the growth and saturation of the Kelvin-Helmholtz instability in a
compressible fluid layer with and without a weak magnetic field. In the absence of a magnetic field, the
instability generates a single eddy which flattens the velocity profile, stabilizing it against further perturbations.
Adding a weak magnetic field - weak in the sense that it has almost no effect on the linear instability - leads to
a complex flow morphology driven by MHD forces and to enhanced broadening of the layer, due to Maxwell stresses. 
We corroborate earlier studies which showed that magnetic fields destroy the large scale
eddy structure through periodic cycles of windup and resistive decay, but we show that the rate of decay decreases with decreasing plasma resistivity $\eta$, at least within the range of $\eta$ accessible to our simulations.  Magnetization increases the efficiency of momentum transport, and the transport increases with decreasing $\eta$.
\end{abstract}

\keywords{instabilities --- methods: numerical --- MHD --- turbulence}

\section{INTRODUCTION}\label{s:introduction}
Turbulent mixing layers, found at the boundary of two fluids in relative
 motion, are thought to  entrain ambient material into outflows
 and to contribute to the near uniform chemically mixed interstellar medium
 (ISM) (see reviews by Scalo \& Elmegreen 2004 and Elmegreen
 \& Scalo 2004).  Shear flow instabilities (ie. Kelvin Helmholtz Instability
 (KHI)) provide the main mechanism of forming these mixing layers.  Because shear flows show up in many places in astrophysics and
 geophysics, the properties of KHI has been studied extensively using both linear stability analysis and non linear simulations (see \S\ref{s:linear}).

In this paper, we expand on these results, focusing primarily on the effects of a weak magnetic field, too
weak to affect the initial development of the instability. A scenario for
the nonlinear evolution of weakly magnetized shear layers was proposed by
Frank et al. 1996 and Malagoli et al. 1996, based on numerical simulations.
The field is wound up in the flow until the point of reconnection,
 which injects energy into smaller scale eddies in the flow.  The field then winds up in these eddies until it reconnects again. This process is continued until the perturbed magnetic energy is damped out,
leaving an enlarged shear layer. It remains to be shown that
 this picture
holds for resistivities $\eta$ appropriate to the ISM, which are much lower than
anything that can be achieved in a numerical simulation. In this paper,
we study the evolution of the layer with $\eta$ varied by a factor of
50. We find that certain properties - notably the time for the instability to saturate, the energy at saturation, and the energy put into the
magnetic field -  appear to saturate with $\eta$. Others - the rate at which perturbations decay and the rate at which the layer broadens with
time - do not reach $\eta$ independent states for the range of $\eta$ accessible in our calculations.

Section \ref{s:problem} describes the basic problem setup and discusses the linear theory. Details on the numerical
method and parameters of our models are given in \S\ref{s:computations}. The results (\S\ref{s:results}) are
summarized in \S\ref{s:conclusion}.

\section{OVERALL PROBLEM}\label{s:problem}
\subsection{Basic Equations \& Equilibrium State}\label{s:equilibrium}
The conservative variables that describe the system are the density $\rho$,
momentum density $\rho\mbfv$, total energy density $\rho E$, and magnetic induction $\mbfB$.  The evolution of these variables is given by the MHD equations.  Written in divergence form, these are:
\begin{equation}\label{e:continuity}
\f{\partial \rho}{\partial t} + \del\cdot (\rho\mbfv) = 0
\end{equation}
\begin{equation}\label{e:momentum}
\f{\partial (\rho\mbfv)}{\partial t} + \del\cdot \left[\rho\mbfv\mbfv -
\f{\mbfB\mbfB}{4\pi} + p + \f{\mbfB^{2}}{8\pi}\right] = \del\cdot\bar{{\bf \Pi}}
\end{equation}
\begin{equation}\label{e:energy}
\f{\partial(\rho E)}{\partial t} + \del\cdot \left[\rho E \mbfv +
  (p+\f{\mbfB^{2}}{8\pi})\mbfv-\f{(\mbfv\cdot\mbfB)\mbfB}{4\pi}\right]=
 \mbfv\cdot\left(\del\cdot\bar{{\bf \Pi}}\right) + \eta\mbfJ^{2}
\end{equation}
\begin{equation}\label{e:induction}
\f{\partial \mbfB}{\partial t} + \del\cdot (\mbfv\mbfB - \mbfB\mbfv) =
\f{\eta c^{2}}{4\pi}\del^{2}\mbfB,
\end{equation}
where we have included terms for both viscous and resistive dissipation. In
the case of an incompressible fluid and a spatially isotropic viscosity, 
the viscous term in the momentum equation reduces to
$\del\cdot\bar{{\bf \Pi}}=\rho\mu\del^{2}\mbfv$, where $\mu$ is the
 coefficient of viscosity.  Numerical schemes cannot resolve the dissipation scales of the ISM,
which is very close to the ideal limit,
 with a Reynolds number, $Re\equiv \f{VL}{\mu}\sim 10^{10}$ and a magnetic
Reynolds number, $R_m\equiv \f{VL}{\eta}\sim 10^{16}$.  We will discuss our treatment of viscous and resistive dissipation in \S~\ref{s:numerics}.
The gas pressure $p$ is related to the energy density by
 $p=(\gamma-1)(\rho E-\f{1}{2}\rho\mbfv^{2}-\f{\mbfB^{2}}{8\pi})$ where $\gamma$ is the adiabatic gas constant,
which we keep fixed at 5/3.

The equilibrium flow is aligned along the $\hat{{\bf x}}$ direction and is sheared in the $\hat{{\bf y}}$ direction,
\begin{equation}\label{e:profile}
\mbfv_{0}=\hat{{\bf x}} \f{V_{0}}{2} \tanh (\f{y}{a}),
\end{equation}
while the gas pressure and density are initially constant, $p=p_{0}$,
 $\rho=\rho_{0}$.  Eq. \ref{e:profile} describes a flow that is in
 the positive $\hat{{\bf x}}$ direction for $y>0$ and in the negative
 $\hat{{\bf x}}$ direction for $y<0$ and smoothly varies within a
shear layer of total width $2a$, asymptotically reaching $\pm\f{V_{0}}{2}$ for
 $\left|\f{y}{a}\right|\gg 1$.  In the absence of
 perturbations, the velocity profile will broaden due to viscous dissipation
 on a timescale $t_{visc}$.  In practice, we restrict
 ourselves to timescales $t$ such that $t\ll t_{visc}$. 
In the MHD case, we take the equilibrium field to be spatially constant
and aligned with the flow:
\begin{equation}\label{e:mag}
\mbfB_{0}=\hat{{\bf x}}B_{0}.
\end{equation}

It will be convenient to describe the problem with dimensionless parameters.  The ratio of
the flow speed and the sound speed, $c_{s,0}=\sqrt{ \f{\gamma
    p_{0}}{\rho_{0}}}$, is the sound Mach number, $M_{s}=\f{V_{0}}{c_{s,0}}$.
 The ratio of the flow speed to the Alfven speed,
 $c_{a,0}=\f{B_{0}}{\sqrt{4\pi\rho_{0}}}$, is the Alfven Mach number,
 $M_{a}=\f{V_{0}}
{c_{a,0}}$.  The strength of the magnetic field can be
 described by the ratio of the Alfven speed to the sound speed,
 $\alpha=\f{c_{a,0}}{c_{s,0}}=\f{M_{s}}{M_{a}}$. The level of viscous
diffusion is characterized by the Reynolds number, $Re=\f{V_0 L}{\mu}$ and that of
resistive diffusion by the magnetic Reynolds number $R_m=\f{V_0 L}{\eta}$, where L is the characteristic length of the system.

\subsection{Linear Stability}\label{s:linear}
 A linear stability analysis will tell us under what conditions an instability
 will arise and how fast we expect it to develop.  A
 linear stability analysis involves perturbing the linearised MHD equations
 and solving for the growthrate $\Gamma$ of the instability
as a function of the perturbed wavenumber, $\mbfk$.  The linear properties of
 shear flow instabilities have been extensively studied
 elsewhere.  What follows is a summary of salient results.

%\subsubsection{Incompressible, Inviscid Vortex Sheet}
Chandrasekhar (1961) solved the incompressible, inviscid, ideal vortex sheet
 problem where the velocity profile is characterized by a
 discontinuity:
$\mbfv_{0}=\hat{{\bf x}}\f{V_{0}}{2}$ for $y>0$, $\mbfv_{0}-\hat{{\bf x}}\f{V_{0}}{2}$ for $y<0$.  In the purely hydrodynamic case
 $(\mbfB_{0}=0)$, assuming $\rho$ is the same in the two media,
 the growthrate is $\Gamma_{HD}=\left|\mbfk\cdot \f{V_{0}}{2}\right|$.  When
 the flow velocity and wave vector line up, all modes are
 unstable with the smallest scales growing the fastest.

A magnetic field that is oriented perpendicularly to the flow plane has no effect on the stability of the
flow.  It only acts to make the fluid more incompressible.  However, a magnetic field aligned parallel to
the flow has a stabilizing effect.  The tension in the magnetic field tends to suppress the instability, modifying the growthrate to:
\begin{equation}\label{e:growthrate}
\Gamma_{MHD}=\Gamma_{HD}
\left[1-\left(\f{2}{M_{a}}\f{\hat{\mathbf{e}}_{k}\cdot\hat{\mathbf{e}}_{B_{0}}}  {\hat{\mathbf{e}}_{k}\cdot\hat{\mathbf{e}}_{U_{0}}}\right)^{2}\right]^{1/2},
\end{equation}
If the initial velocity, magnetic field, and wave vector all line up, the condition for stability is $M_{a}\equiv\f{V_{0}}{c_{a,0}}<2$.  In other words, the Alfven speed has to be larger than half the shear velocity
difference($c_{a,0}>\f{V_{0}}{2}$) to stabilize the layer.

%\subsubsection{Effect of a Shear Layer}
The effects of a shear layer, from a linear profile (eg. Ray 1982) to a
 $\tanh$ profile (Lau \& Li 1980, Miura \& Pritchett 1982) have also been
 looked at.
 Ferrari \& Trussoni (1983) summarized the previous work by examining profiles that range between these two cases.
 The effects of different profiles
are all qualitatively the same, though the exact details depend on the
 initial equilibrium.  In general, introducing a layer adds another lengthscale
 to the problem.  For wavelengths that are long compared to the size of the
 layer, the details of the layer become unimportant and the growthrate
 approaches
 the vortex sheet result.  However, when the wavelength is short compared to the layer width,
$1/\mbfk \ll a$, the perturbation no longer "sees" the shear and becomes
 stable.  As a result, the fastest growing modes are no longer the shortest
 wavelengths
 but are now ones whose wavelength is comparable to the size of the layer,
 $\mbfk a \sim 1$.  A parallel magnetic field still tends to stabilize the
 flow,
 though the stability criterion ($M_{a}>2$) only holds
approximately.  The exact criterion depends on the details of the problem. 

%\subsubsection{Effect of Compressibility}
The effect of compressibility on shear flows has been studied by Landau
 (1944), Miles (1957), Fejer (1963) and others for a vortex sheet, and by
 Blumen (1970),
 Blumen, Drazin, \& Billings (1975), Drazin \& Davey (1977), Ray (1982), and
 Ferrari \& Trussoni (1983) for a shear layer.  Compressibility has the
 general
 effect of lowering the growthrate from the incompressible case.  Also, Miles
 (1957) showed that compressibility adds an upper Mach number limit on the
 stability
 for a vortex sheet for parallel perturbations (it remains unstable to 3D perturbations): for $M_{s}>\sqrt{2}$ the flow becomes stable.  However, in the presence of a shear layer, all Mach numbers are unstable, but
the growthrates in the supersonic case are about an order of magnitude less
than in the  subsonic case.  The spatial distribution of mode energy is
sensitive to $M_s$ as well:  for  subsonic flow the  modes decay 
exponentially away from the layer, while in supersonic flow
they decay much more slowly (Blumen et al. 1975). 

Hughes \& Tobias (2001) summarize and expand the literature regarding the
 differences between a 2D and 3D treatment of the linear theory of shear
 flow instabilities.   In a purely hydrodynamic flow, for every unstable mode in 3D, there is a corresponding unstable mode in 2D at a lower Reynolds number,
$Re\equiv aV_0/\mu $. Therefore, as $Re$ is increased to a value that is no
longer stabilizing, the 2D mode will become unstable first (Squires 1933) 
As a consequence, a 2D stability analysis
of hydrodynamic flows is adequate.  Michael (1953)and Stuart (1954) showed that, for an MHD flow, for every unstable 3D mode there is a corresponding 2D mode at both a lower Reynolds number $Re$ and magnetic Reynolds number $R_m$.  In this case, the region of neutral stability is a function of both $Re$ and $R_m$.  Therefore, the 2D mode will not necessarily become unstable first (Hunt 1966).  In the limit of ideal MHD ($Re=\infty$ and $R_m=\infty$) an unstable 2D mode is a faster growing mode than the corresponding 3D mode (Hughes \& Tobias 2001).  Therefore, although it is adequate to study ideal MHD flows in 2D, non-ideal MHD flows may have to be treated in 3D. 

In order to develop some feeling for the effects of resistivity and
viscosity, and for the effect of a 3D wavevector for the set of models we ran, we carried out a parameter study using a linear eigenvalue code.  Our code solves the linearized incompressible MHD equations with respect to a background flow and magnetic field.  The linear perturbation is Fourier decomposed in the $\hat{{\bf x}}$ and $\hat{{\bf z}}$ directions, and the remaining $\hat{{\bf y}}$ direction is discretized with a Chebyshev spectral method (Boyd 2001, Trefethen 2000).  The eigenvalue problem is solved in an infinite domain $y\in [-\infty,\infty]$, which is mapped into $y'\in [-1,1]$ by a mapping scheme that localizes most of the grid points near $y=0$.  This ensures that we resolve the instability, and is compatible with the boundary conditions that all perturbations vanish as $y$ goes to $\pm\infty$.  We used 150 grid points along $y$ to perform a parameter scan in viscosity and resistivity to get a handle on the relative effects on both.  We have also looked at the effect of an increasing $k_{z}$ component to the perturbation.  In both studies, we have used the equilibrium flow described in \S~\ref{s:equilibrium} and initial parameters given in \S~\ref{s:models}. 

\begin{figure}
\centering
\includegraphics[width=3in]{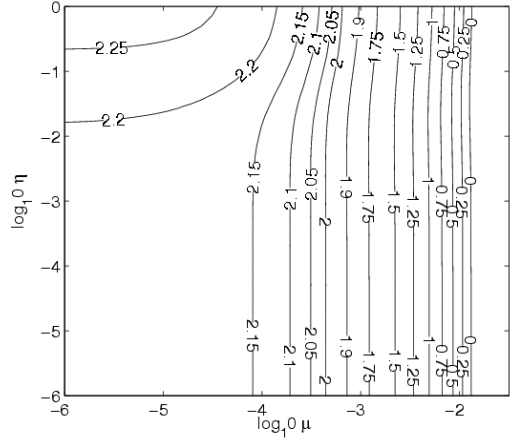}
\caption{Contours of growthrate are plotted as a function of the viscous, $\eta$, and resistive, $\mu$, diffusivities for a flow speed and magnetic field comparable to those given in Table \ref{t:models}.  Viscosity acts to stabilize the instability by dissipating kinetic energy while the resistivity acts to destabilize the instability by decoupling the field from the flow.  At low viscosities, the growthrate only varies by a few percent because we are considering only a weak magnetic field.}
\label{fig:parascan}
\end{figure}

Figure \ref{fig:parascan} shows contours of growthrate as a function of both the viscous, ($\mu$), and resistive, ($\eta$), diffusivity.  As long as the viscous diffusivity is $\mu \gtrsim 0.0003$, the growth rate is insensitive to resistivity and decreases with increasing viscosity, eventually becoming stable.  At low viscous diffusivity ($\mu \lesssim 0.0003$) the effect of viscosity becomes negligible. Resistivity only starts to  make a non-negligible difference for $\eta \gtrsim 0.001$, above which it acts as a destabilizing mechanism.  These two observations make sense because viscosity acts as a dissipation mechanism for the kinetic energy of the instability and resistivity, though it is a dissipation mechanism for the magnetic energy, also affects the frozen-in level of the magnetic field, which is a stabilizing mechanism.  As $\eta$ increases the field decouples from the fluid and the growthrate approaches the hydrodynamic growthrate.  As $\eta$ decreases, the field becomes more frozen-in and will then approach the ideal MHD growthrate.  Since we are considering a weak field case only, the amount of stabilization provided by the magnetic field is small and therefore the differences between the MHD and hydro growthrate is small.

\begin{figure}
\centering
\includegraphics[width=3in]{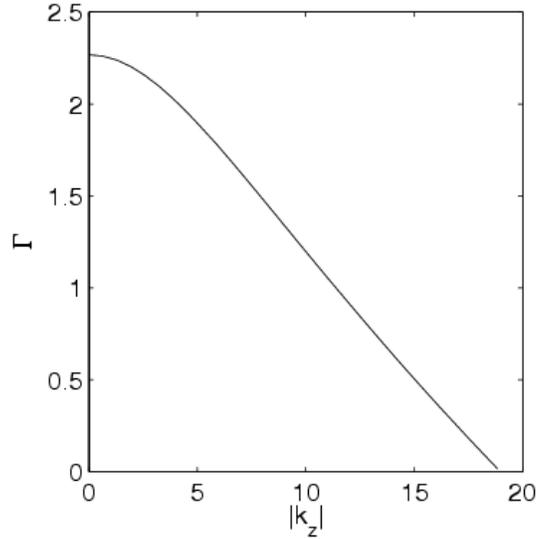}
\caption{The growthrate as a function of $k_{z}$ is plotted for flow speeds and magnetic fields comparable to those given in Table \ref{t:models}.  The fastest growing mode for the parameters considered is the $k_{z}=0$ mode.  As a consequence a 2D study should be an adequate place to start.}
\label{fig:kzscan}
\end{figure}

Fig. \ref{fig:kzscan} shows the growthrate as a function of $k_{z}$ with a constant $k_{x}$ for similar flow properties as given in Table~\ref{t:models}.  The fastest growing mode occurs when $k_{z}=0$ and decreases for growing $k_{z}$ until it becomes stable.  Since we are considering a weak-field problem in which resistivity and viscosity have little effect on the growthrate (see Fig.\ref{fig:parascan}), it is not surprising that the 2D mode is the most unstable.

%\subsection{Non-Linear Analysis}

\section{THE COMPUTATIONS}\label{s:computations}
\subsection{Models Run}\label{s:models}
We focus on the non-linear evolution of the Kelvin-Helmholtz instability in weak-field MHD flow.  For realistic values of viscosity and resistivity, the instability growthrates
in  the two flows are similar (see Fig. \ref{fig:parascan}).  However, the nonlinear evolution of the two are very different (Frank et al. 1996, Malagoli et al. 1996). As explained in the Introduction, we wish to examine the role of resistivity in the evolved flow, and to compare momentum transport in the HD and weak field MHD flows.

Table \ref{t:models} provides a summary of the models run. The numbers in the names of the MHD runs refer to the magnetic Reynolds number $R_m$.
Even though the maximum value $R_m=50000$ is far below that of the ISM ($R_m\sim 10^{16}$) we are able to draw useful conclusions (see \S 4.3).
 The entries in
the last column denote "low" (256x512), "medium" (512x1024), "high"
(1024x2048), and "super high" (2048x4096) resolution, respectively; the multiple runs being used to test
convergence. All the runs have the same initial pressure ($p_{0}=
1.0$), density ($\rho_{0}=1.0$), sonic Mach number $M_{s}=1$ and the
MHD runs all have  $M_{a}=10$, so $\alpha=0.1$ and $\beta=120$.  
We solve eqns. (1) - (4), with the equilibrium flow given by eqn. \ref{e:profile} and eqn. \ref{e:mag}, and an initial velocity perturbation  given by:
\begin{equation}\label{e:perturbation}
\delta\mbfv_{y}=\delta V_{0}\sin(k_{x} x) \exp(-\f{y^{2}}{\sigma^{2}})
\end{equation}
where the amplitude $\delta V_{0} \ll \f{V_{0}}{2}$. The  attenuation with distance from the layer on a lengthscale given by $\sigma$ ensures that the perturbation stays localized to the $y=0$ interface.  

The computations are done on a 2D grid in the x-y plane.  The dimensions of the grid are $x=
[0,L]$ and $y=[-L,L]$ where L is the characteristic length of the system.  We choose the characteristic length such that a single wavelength fits in the box, $L=2\pi/k_{x}$.  We have chosen $k_{x}=2\pi$ such that the growthrate is near a maximum (see Keppens et al. 2001), and we set $a/L=0.05$ and $\sigma /L=0.2$.  These are the same initial conditions used by both Malagoli et al. (1996) and Keppens et al. 2001).  These choices allow a small, yet resolvable shear layer, and an attenuation scale that is both larger than the layer ($\sigma /a=4$) and shorter than the characteristic length, so that the $\pm y$ boundaries interact minimally with the instability. We have chosen the characteristic time to be the sound crossing time $t_{s}=L/c_{s,0}$.  All the models presented here are run over a timescale of $20t_{s}$, which is long enough to capture the saturation and at least the initial non-linear evolution of the instability.

\subsection{Numerical Method}\label{s:numerics}

The magnetohydrodynamical scheme -- Proteus -- 
is based on a conservative gas-kinetic flux-splitting
method, introduced by Xu~(1999) and Tang \& Xu~(2000).
Viscosity and ohmic resistivity (RHS of eq.~\ref{e:induction}) are 
implemented via dissipative fluxes (Heitsch et al. 2004, 2007). 
Thus, Proteus allows the full control of dissipative effects. Test
cases have been presented by Heitsch et al.~(2007). Proteus
integrates equations~(\ref{e:continuity}-\ref{e:induction}) at
2nd order in time and space. We use only the resistivity
implementation for the models presented here, relying on the (extremely
low) numerical viscosity. 

Proteus is also equipped with a Lagrangian tracer mechanism (Heitsch et al. 2006), 
which permits us to follow particles advected by the flow and thus enables us
to study mixing. We defer a discussion of the mixing properties of the flow to a 
future paper (Palotti et al. 2007) except for a brief mention in \S 4.1.

\section{RESULTS}\label{s:results}
\subsection{Convergence}\label{s:convergence}
The numerical dissipation scale is related to the resolution 
(although usually not directly analytically to $\Delta x^2/\Delta t$). As
resolution increases, the dissipation scale decreases, and structure
is allowed to develop on those smaller scales.

Setting the dissipative scales allows us to reach detailed convergence (i.e. a
``point-to-point'' agreement in physical variables between resolutions) in 
the linear regime of the instability. Detailed convergence usually cannot be reached
in the non-linear (or turbulent) regime, however, virtual convergence (i.e.
the convergence in integrated quantities between resolutions, such as magnetic energy)
is still possible if the problem is nearly independent of the dissipation scales.
We will be using ``convergence'' in the latter, virtual, sense.

In a weak field MHD flow, the magnetic field becomes wound up in the flow,
increasing the current density and enhancing the role of resistive
dissipation. As we show in the ensuing discussion, resistive relaxation is
rather abrupt\footnote{We do not have the
resolution to see whether the flow resembles any of the standard modes
of reconnection, such as Sweet-Parker reconnection, but it has been shown
elsewhere that shear flow instability can enhance the reconnection rate
(Knoll \& Chac\'on 2002).}. 
The resistive event alters the structure of the flow.  Therefore, resistivity plays an integral part in determining the evolution of the system.  Viscosity, on the other hand, becomes important when energy cascades down to the dissipation scale and appears to be unimportant in determining the bulk properties of the flow.  For these reasons, we use a physical resistivity while relying on numerical viscosity. 

We have carried out a convergence study by running each of our models (see Table \ref{t:models} at
resolutions ranging from 256$\times$512 to 2048$\times$4096.  In Fig. \ref{fig:convergence} we present the (virtual) convergence of the total kinetic energy in the y direction, $KE_{y}=\int\int\f{1}{2}\rho v_{y}^{2}dxdy$, as a function of time for the HD model, as well as the total perturbed magnetic energy, $\Delta ME=\int\int\f{B^{2}-B_0^2}{8\pi}dxdy$, for the models MHD5000 and MHD50000.  

Any discrepancies in the different resolutions for the $KE_{y}$ in the HD model can be attributed to numerical viscosity only.  It appears that $KE_{y}$ is a converged quantity.  Even though there is not true convergence, the differences between the 512 $\times$ 1024 and 1024$\times$ 2048 models are less than 5\% while those of the 256$\times$512 and 512$\times$ 1024 can be as high as 10\%, ie. the differences are small and decrease with increasing resolution.

The perturbed magnetic energy is a good indicator of our ability to resolve the resistive length scales.  For the larger resistivity, $R_m=5000$, is well converged.  For times $t<16t_{s}$ the difference between the 512$\times$1024 and 1024$\times$2048 runs is less than $2\%$ and reaches at most $\sim$ 5.5\%.  Achieving convergence in the MHD50000 model is much more difficult.  For times $t<14t_s$ the differences between the 1024$\times$2048 and 2048$\times$4096 runs are less than $\sim$ 5\%.  At later times, the differences reach up to $\sim$ 15\%.  This is an indication that we have not fully resolved the resistive length scale.  However, the rate of dissipation of the magnetic energy qualitatively follows the same trend for both the 1024$\times$2048 and 2048$\times$4096 runs.  Therefore, for $t<14t_s$, which includes both the saturation of kinetic energy ($t=6.4t_s$) and the saturation of the magnetic energy ($t=9.0t_s$), all the runs are effectively converged and for later times, runs with $R_m\lesssim5000$ are effectively converged while those with lower resistivity only show qualitatively similar trends.

%Achieving convergence in the lower resistivity model is much more difficult.  Even so, for times $t <16 t_{s}$ the differences are low and at later times the trends are qualitatively the same.

\begin{figure}
\centering
\includegraphics[width=3.5in,angle=90]{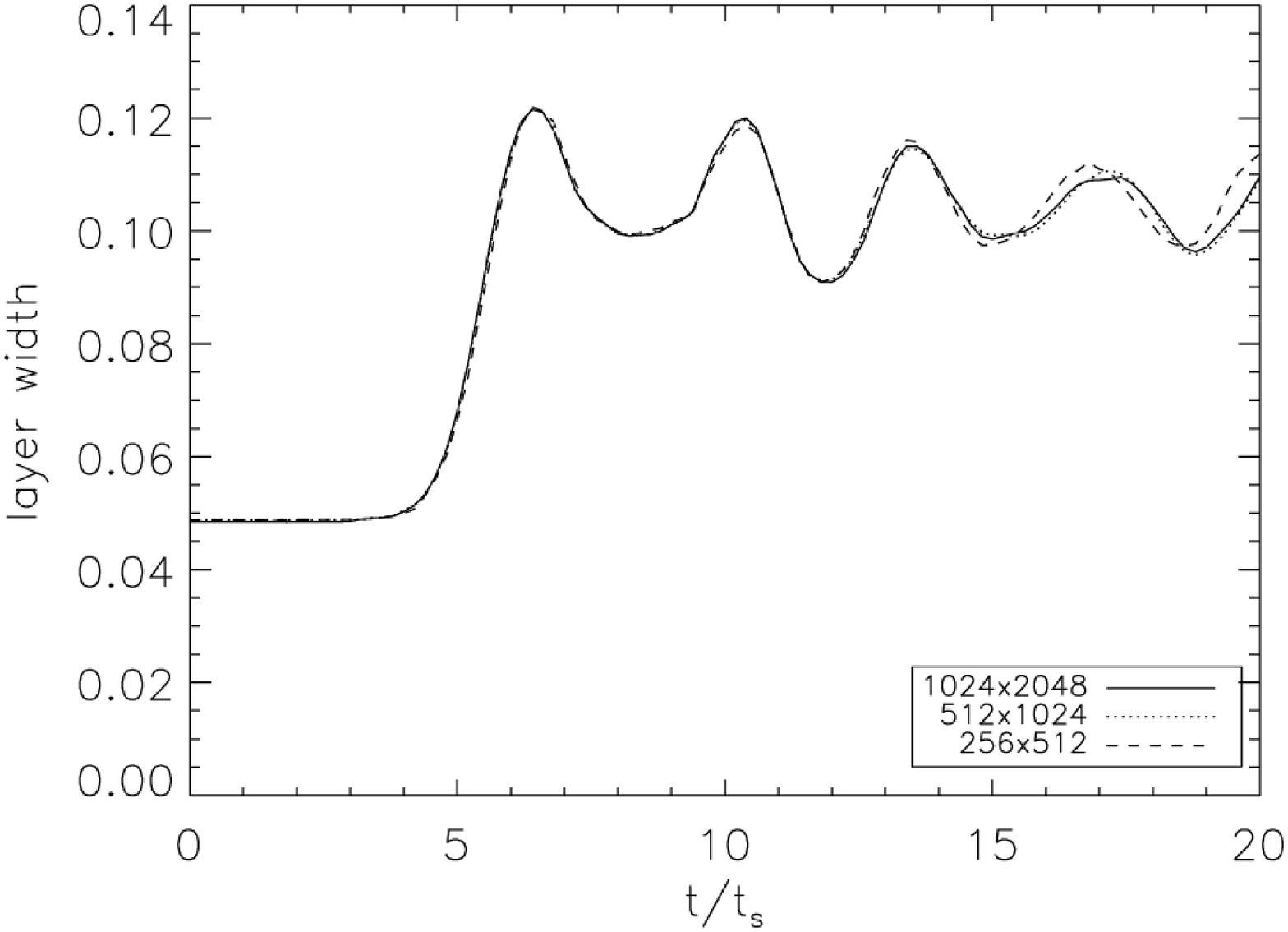} \\
\includegraphics[width=3.5in,angle=90]{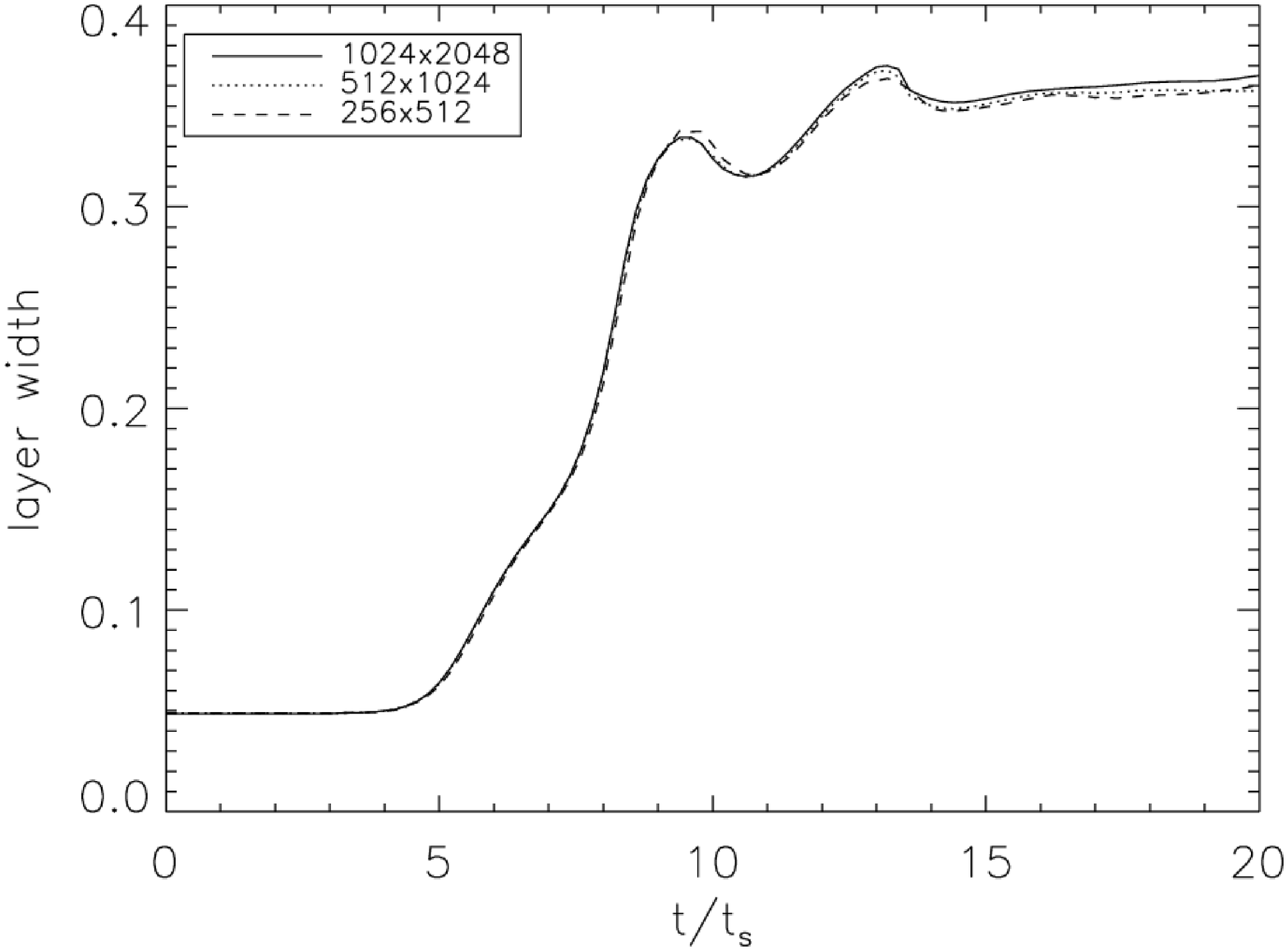}
\caption{The width of the velocity profile as a function of time for both the HD (top) and MHD5000 (bottom) models.}
\label{fig:profilewidth}
\end{figure}

As discussed in \S 2.1, we chose the initial horizontal velocity $v_x$ to be proportional to $\tanh{(y/a)}$ (see eqn. (\ref{e:profile})), with
the parameter $a$, which we call the layer width, set to $0.05$. If we take horizontal averages of $v_x$ at constant $y$ at later times,(see Fig.~\ref{fig:profile} for examples), the transition region between $V_0$ and $-V_0$ is well fit by a straight line.  From this we can define the width of the layer as the position where the velocity, $v=\f{V_0}{\sqrt 2}$.   We find the layer width is an increasing function of time.  
In Figure \ref{fig:profilewidth}  we present the velocity layer width as a function of time for the  HD run and the MHD run MHD5000.  Over the entire time run, both layer profiles agree to within a few percent for the highest resolution runs. This, together with the convergence of $KE_y$ and $\Delta ME$, is good evidence that
the global properties of the flow are well converged.

As we mentioned in \S 3.2, we can trace the trajectories of tracer particles (for the higher resolution runs, however, this requires a prohibitively
large amount of computer time).  In
order to test the convergence of the small scale structure, we computed the separation of pairs of particles as a function of time for the run MHD1000.  The two highest resolution runs agree for times less than $15t_{s}$.  The differences after this can be attributed to the effects of numerical viscosity on the flow. We defer other discussion of the tracer particles to Palotti et al. (2007).

For timescales $t\lesssim16t_{s}$, each of the models can be considered converged.  The lack of convergence at longer times is a result of the development of small scale flow structure and consequent viscous dissipation, which is entirely numerical.  As the resistivity increases, less small scale structure develops, and thus the convergence is better. At the lowest resistivities, numerical resistivity may be becoming important as well.

\subsection{Nonlinear Flow Morphology}\label{s:morphology}
The growth rates in the HD and MHD models differ very little, because the initial magnetic field is very weak (see Table 1).
Over the range of resistivities we examined ($R_{m}=1000-50000$),
we calculated the growth rates to differ by $\sim$1\%, increasing as the 
resistivity increases. The trend reflects the progressively weaker coupling
of the field to the fluid with increasing resistivity, and hence a reduction of the stabilizing magnetic tension force.

However, the nonlinear evolution of these flows are
vastly different.  In the HD flow, a single large eddy develops and remains until viscosity eventually
damps it out, on timescales much longer than considered here.  The total kinetic energy in the y-direction, the top panel in Fig. \ref{fig:convergence}, is consistent with a tumbling elliptical eddy.  The amplitude of each successive oscillation decreases due to viscosity.  The density structure and velocity field at time $t=8.4t_{s}$ for the HD model are plotted in Fig. \ref{fig:hdmorph}.  The density shows a rarefied central region around which the single eddy spins.  The density contrast between the central region and the outer region is $\sim$ 36\% in this case.
The tumbling eddy is also evident in the velocity field (see Fig.\ref{fig:hdmorph}).
  
The evolution of the MHD model is much more complicated, and certain aspects of it depend on resistivity.  We have plotted the magnetic energy structure (grey scale) and magnetic field (vectors) for three times, $6.4t_{s}$,$12t_{s}$, and $20t_{s}$, and two values of $R_m$, 5000 and 50000, in Figure (\ref{fig:mhdme}). These times represent the saturation of $KE_y$ ($6.4t_s$), the end of the run ($20t_s$) where, at $R_m \lesssim 5000$, both $KE_y$ and $\Delta ME$ have nearly decayed fully, and an intermediate time ($12t_s$) where the energies have partially decayed.  Our primary observations about Fig. (\ref{fig:mhdme}) are that (1) the magnetic field is wound up by the large scale eddy, and that this wound up field induces a wealth of magnetic
 structure, that (2) reducing the resistivity allows the structure to exist at smaller scales, and
that (3) reducing the resistivity allows the structure to persist for longer times.  
Some of the magnetic folds and loops seem to be magnetic islands, which can only
form through resistive evolution. Whether this is better described as reconnection or diffusion is unclear, however (see footnote 1).  Along with the magnetic structure, there is a corresponding small-scale density and velocity structure in the MHD runs, with density perturbations about $\pm$ 2\% for $R_m=5000$ and a maximum of about 2\% and minimum of about 9\% for $R_m=50000$.

\subsection{Saturation and Effect of Resistivity}\label{s:saturation}
\begin{figure}
\centering
\includegraphics[width=3.5in,angle=90]{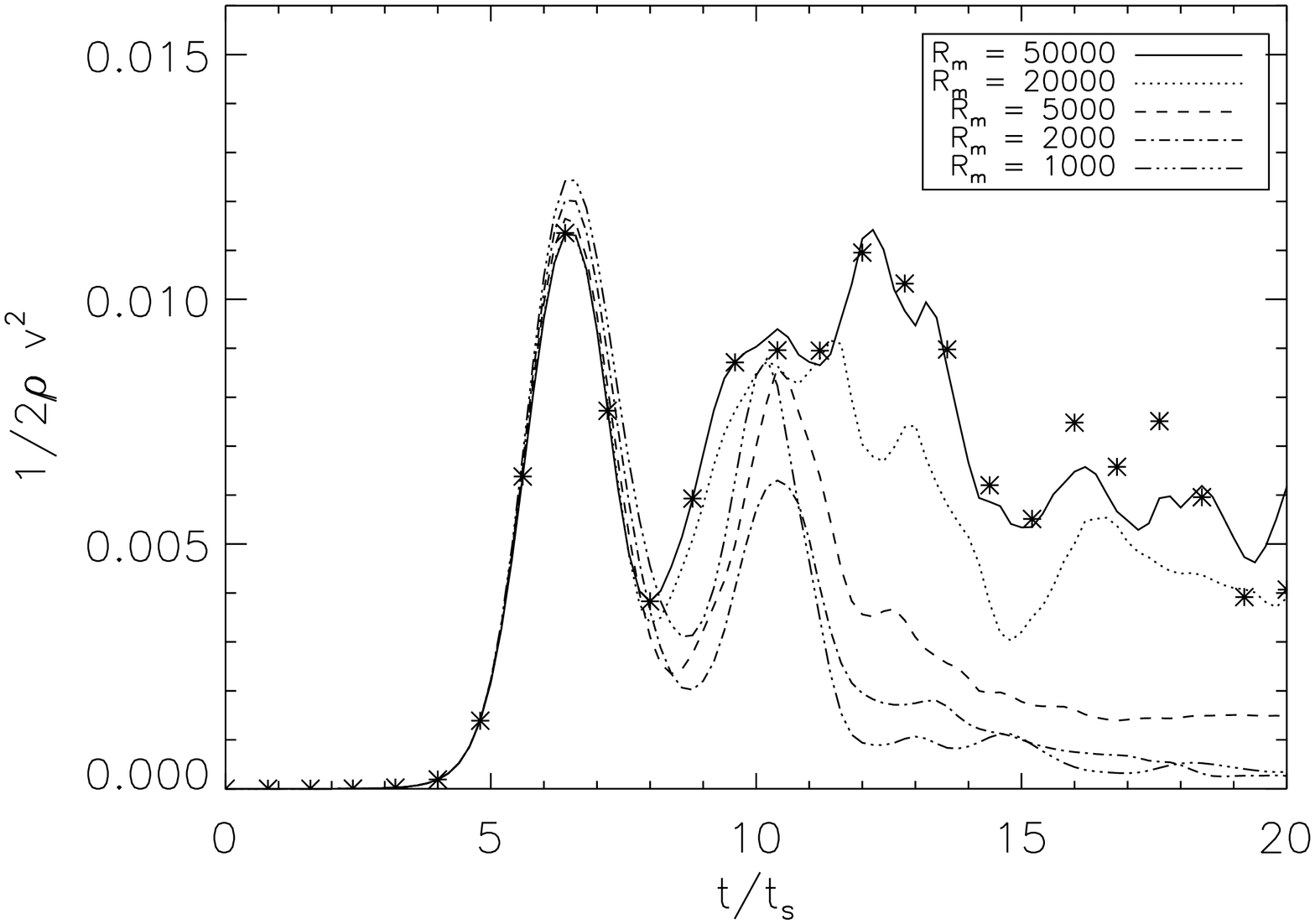} \\
\includegraphics[width=3.5in,angle=90]{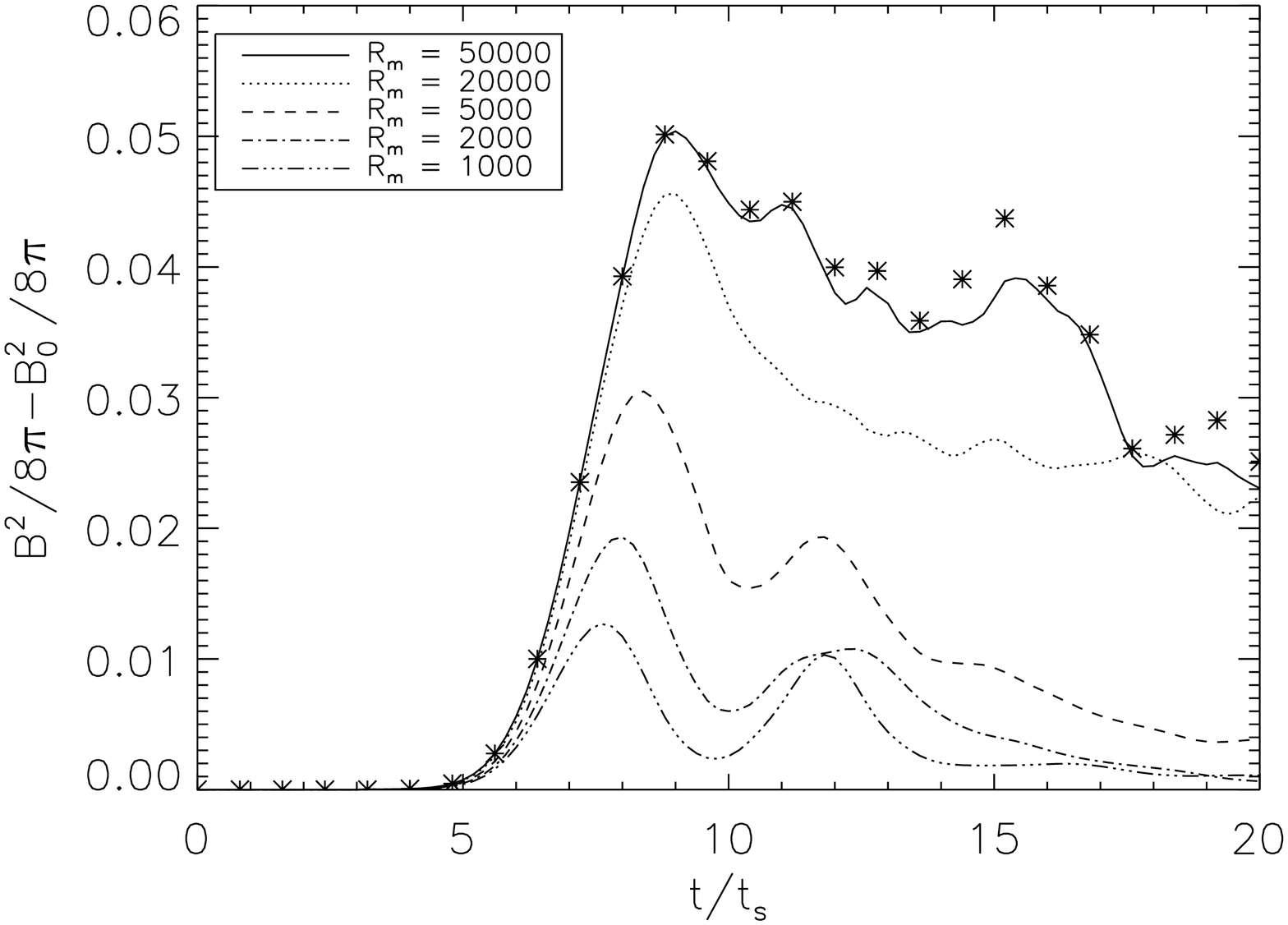}
\caption{ The  $y$-component of the kinetic energy, $KE_y$ (top) and the perturbed magnetic energy $\Delta ME$ (bottom) as a function of time for magnetic Reynolds number ranging from $R_m=1000$ to $R_m=50000$.  Because the $R_m=50000$ run is not as well converged, we have also included the highest resolution data as the asterisks.}
\label{fig:energyall}
\end{figure}

Saturation of the KHI is defined as the point at which the kinetic energy in the y direction peaks in magnitude.  We found the time of kinetic energy saturation $t_K$ in both the HD and all MHD runs to be $6.4t_{s}$ (see top panel of Fig.~\ref{fig:convergence} and Fig.~\ref{fig:energyall}). The mechanism for saturation is the interplay between the y components of the forces on the plasma, namely gas pressure, Reynolds stress and Maxwell stress. 
Integrating the y component of the momentum equation (eq.\ref{e:momentum}) from $x=0$ to $x=L$ and from $y=0$ to $y=L$ (the upper half of the domain) and applying periodicity in $x$, we get
\begin{equation}
 \f{\partial \langle\rho v_{y}\rangle}{\partial t} = - \int_{0}^{L} \left(\rho v_{y}^{2} +p+ \f{(B_{x}^{2}-B_{y}^{2})}{8\pi}\right)dx\bigg|_{y=0}^{y=L},
\end{equation}
where we have assumed the viscosity to be negligible.  When $\partial \langle\rho
 v_{y}\rangle/\partial t=0$ the y-forces balance and the instability is
 saturated. We have plotted the the various contributors to the y component of the the
  force for our MHD50000 and HD runs in Fig.\ref{fig:saturation}.  The Reynolds stress
term ($-\rho v_{y}^{2}$) is the component of force that drives the instability
  while the gas pressure ($-p$) component acts to slow down the instability in
the HD model.  In the MHD models, the Maxwell stress term ($\f{B_{y}^{2}}{8\pi}$) also acts to slow down the instability initially, but it is clear
from  Fig.\ref{fig:saturation} that magnetic forces are much smaller in
magnitude than hydrodynamic forces in this flow.

\begin{figure}
\centering
\includegraphics[width=4in,angle=90]{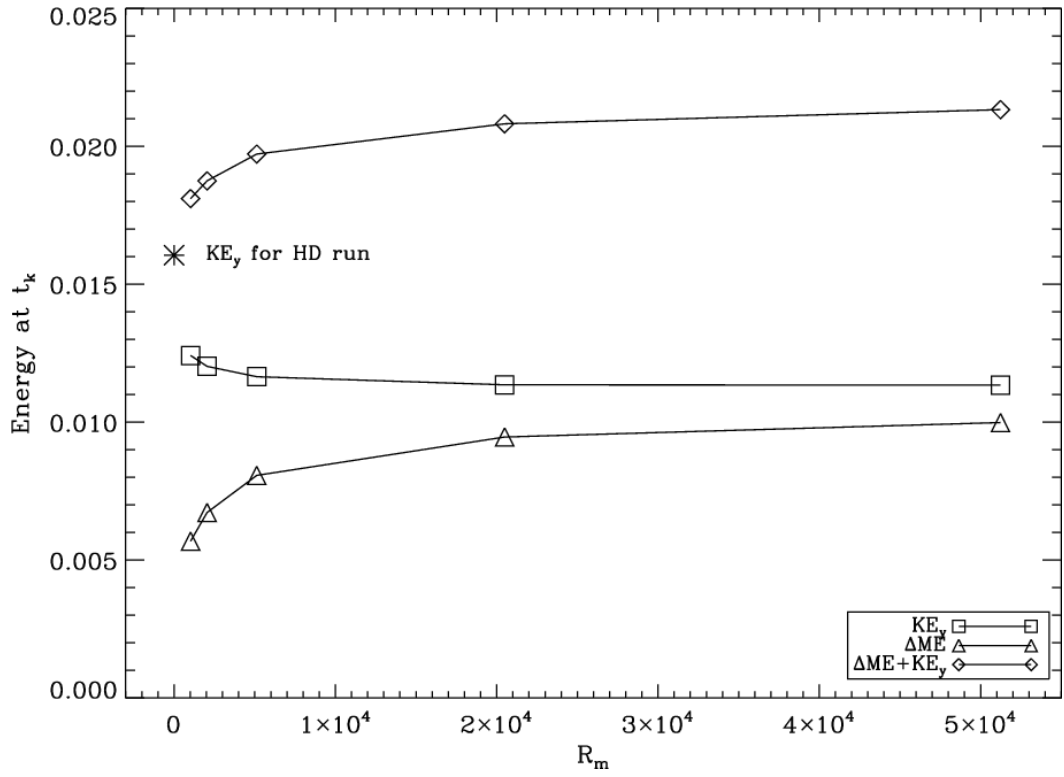}
\caption{The total y kinetic energy, perturbed magnetic energy and total perturbed energy at saturation as a function of resistivity.}
\label{fig:satres}
\end{figure}

We also examined how resistivity affects the saturation. In
 Fig.~\ref{fig:satres} we plot the kinetic energy in the y direction, the total perturbed magnetic field energy, and the sum of the two at $t_K$
as a function of magnetic Reynolds number.  For the sake of comparison,
we have included the HD results at $R_{m}=0$, assuming that as the resistivity becomes infinite, the MHD flow approaches the HD limit.  At $t_K$, the kinetic energy is greatest for the HD model and decreases for increasing $R_{m}$.  The magnetic energy, on the other hand increases for increasing magnetic Reynolds number so that the total perturbed energy increases with $R_{m}$.  By the time of saturation, the magnetic energy has increased between 35\% (lowest $R_{m}$) to 48\% (highest $R_{m}$) over the background value.  As $R_{m}$ increases ($R_{m}>5000)$, the perturbed kinetic and magnetic energies seem to approach equipartition,
 with a ratio of $KE_{y}/\Delta ME_{tot} = \sim 1.14$, whereas at $R_{m}<5000$, the ratio is 2.8.

\begin{figure}
\centering
\includegraphics[width=3.5in,angle=90]{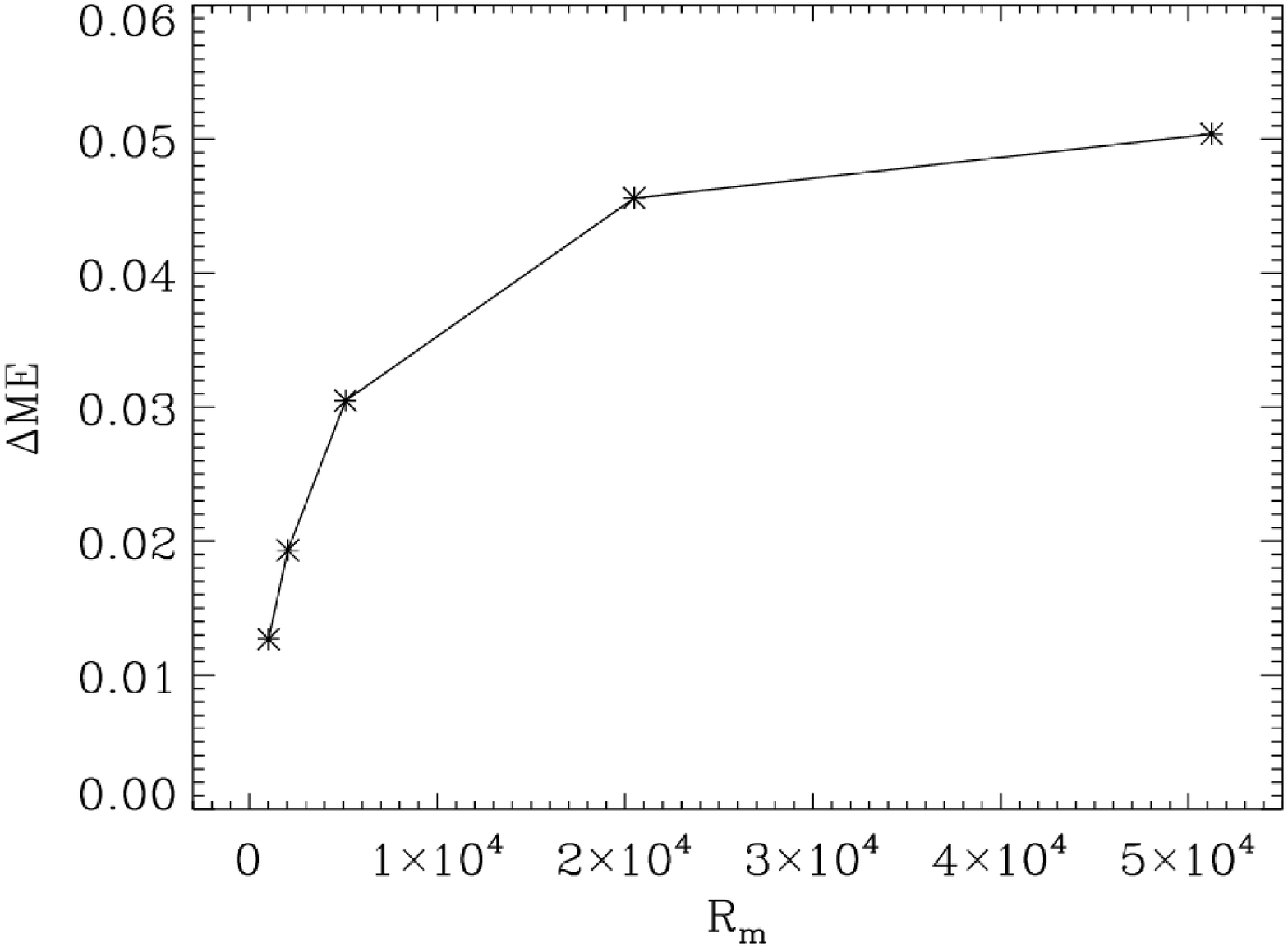} \\
\includegraphics[width=3.5in,angle=90]{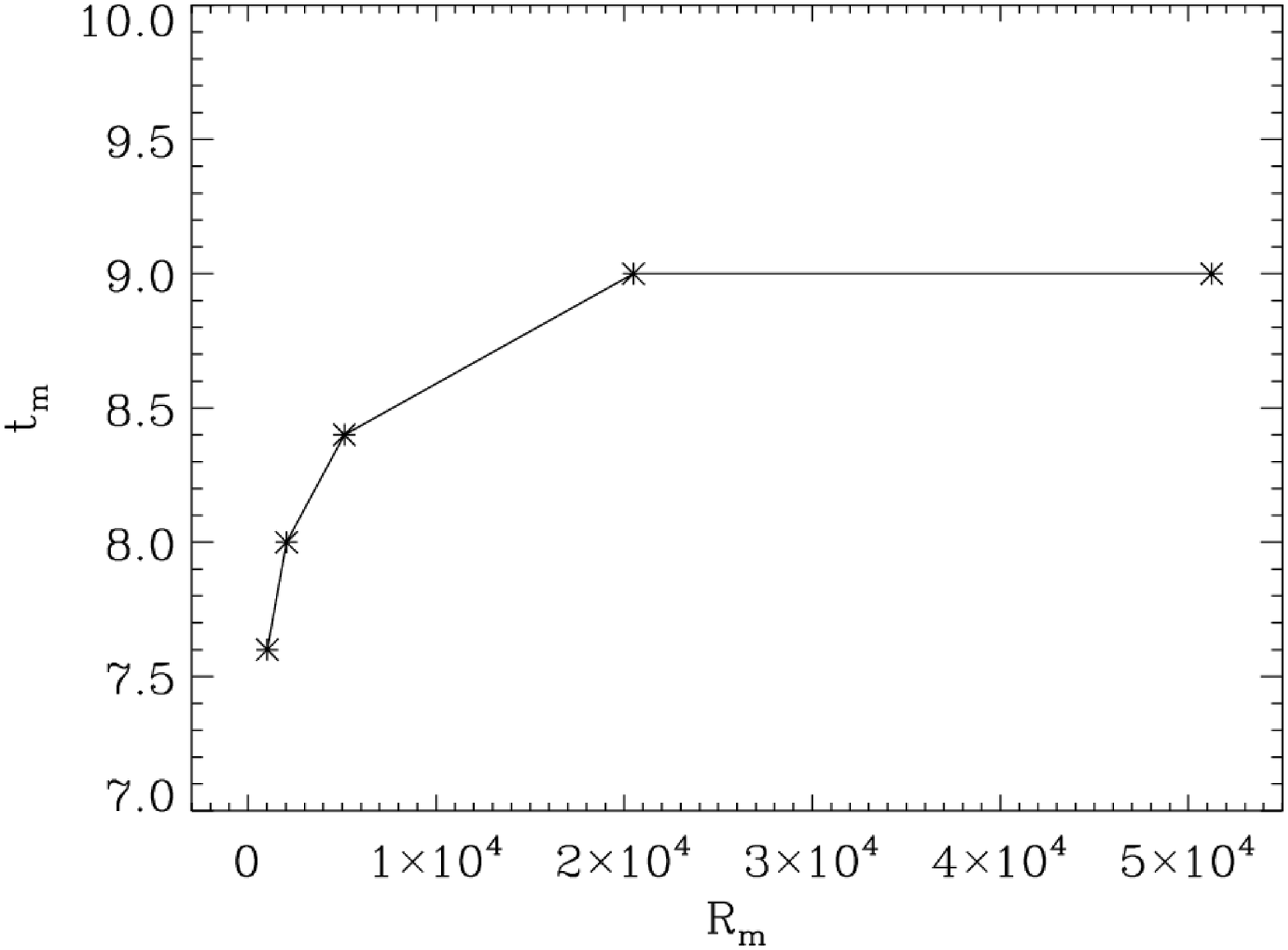}
\caption{The maximum perturbed magnetic energy (top) and the time to reach the maximum perturbed magnetic energy (bottom) as a function of resistivity.}
\label{fig:memax}
\end{figure}

After the kinetic energy saturates, the magnetic energy continues to grow (see Fig. \ref{fig:mhdenergy}), eventually saturating at a later time,
$t_M$. Both $t_M$ and
the value of the magnetic energy at $t_M$ depend on resistivity.  We plot both of these in Fig. \ref{fig:memax}.  As the magnetic Reynolds number increases, the time at which the magnetic energy saturates plateaus at $t_{m}\sim 9.0$, while the maximum magnetic energy continues to increase. 

In order to get a better handle on the mechanism behind the magnetic
saturation, we calculated the volume integrated time rate of change of the
magnetic energy, ignoring the surface terms,
\begin{equation}
 \f{\partial}{\partial t} \left\langle\f{B^2}{8\pi}\right\rangle =  -\langle\eta \mbfJ^2\rangle - \langle\mbfv\cdot (\mbfJ\times\mbfB)\rangle.
\end{equation}
The first term on the right hand side represents the ohmic dissipation while
the second term represents the work done by field on the gas.  We found that
for all models run, during the saturation and decline of the magnetic energy,
the work term is positive meaning that the flow is doing work on the field.
Therefore, the decline in magnetic energy is not dynamical - the field never becomes strong enough to unwind itself -
but is purely dissipative in nature, at least in a global sense, over the range in $R_m$ considered here. The increases in
$t_M$ and $\Delta ME$ with $R_m$ reflect the fact that as time progresses, the flow puts energy into the field at smaller and smaller scales. 
Roughly speaking, the amplification is quenched when the energy reaches the resistive scale; this happens sooner, and at lower magnetic energy,
for the smaller $R_m$. However, it seems likely that $\Delta ME$ is bounded above by the kinetic energy in the flow. The flattening of 
$\Delta ME$ with $R_m$ seen in Fig. (\ref{fig:satres}) is evidence for this. Since this is a dynamical bound, not a resistive one, we 
speculate that there is a transition from resistive to dynamical saturation at resistivities lower than what we can examine.

\subsection{Time Evolution}\label{s:timeevolution}

Figure (\ref{fig:mhdenergy}) depicts a cycle of growth and decay in both kinetic and magnetic energy for $R_m=5000$ (similar evolution is observed for smaller $R_m$ as well). After the kinetic energy
saturates, the magnetic energy continues to grow. Since the growth is at the expense of kinetic energy, $KE_y$ decreases during
this time. As the field grows its structure becomes more complex. Eventually the rate of work done by the flow is overcome by
resistive dissipation. During the decay of the field the kinetic energy grows again, energized in part by the small scale forces characteristic of magnetic reconnection and in part by the background velocity shear that drives instability in the first place.
The cycle repeats itself, but at lower amplitude due to resistive dissipation. After the second peaks, the perturbations simply
decay. This is the behavior found by Frank et al. (1996) and Malagoli et al. (1996).

Figure (\ref{fig:energyall}) 
shows that the picture is more complex, and depends on $R_m$. At $R_m = $ 20000 and
50000, the decay is much less pronounced\footnote{Since the computations at $R_m=50000$ are not as well converged as for the lower $R_m$ (see Fig.~\ref{fig:convergence}), we have plotted values of the energy for the highest resolved run in this case as asterisks.  Comparison of the asterisks with the curves only underscores the basic point as the decrease in magnetic energy is, if anything, slower at the higher resolution}. Whereas for $R_m\le$ 5000 the perturbations have dissipated almost completely by the end of the run,
for $R_m = $ 50000 the perturbed kinetic and magnetic energy densities are still more than half their maximum values. 
The temporal evolution is less cyclic, and shows bumps
and dips that might be associated with discrete magnetic reconnection or dynamical relaxation events, superimposed on a general decay.

Most importantly, there is no sign that the decay rate has reached an asymptotic limit independent of $R_m$. It seems to us 
that there
are two possibilities for the large $R_m$ behavior. One is that the decay rate is determined by a dissipation coefficient other than resistivity. 
Since the viscous diffusivity exceeds the magnetic diffusivity in most of the ISM, viscosity is a likely candidate. Another possibility,
if the gas is weakly ionized, is ion-neutral friction. We are currently simulating flows in which ion-neutral damping dominates; Palotti 
et al. (2008).
The alternative is that the decay rate is determined by a turbulent diffusivity that is independent of any microscopic diffusion coefficient
and is attained at higher $R_m$ than we can achieve. We have no basis for determining which of
these scenarios would hold in the ISM, where $R_m$ is $\sim$ 11 orders of magnitude
larger than it is in our simulations.

One way or another, the energy in the layer must eventually be dissipated. Does it matter how? We argue that it does, both for the observational
appearance of the layer and for the transport taking place within it. This is demonstrated in the following subsection.

\subsection{Momentum Transport}\label{s:momtransport}
\begin{figure}
\centering
\includegraphics[width=3.5in,angle=90]{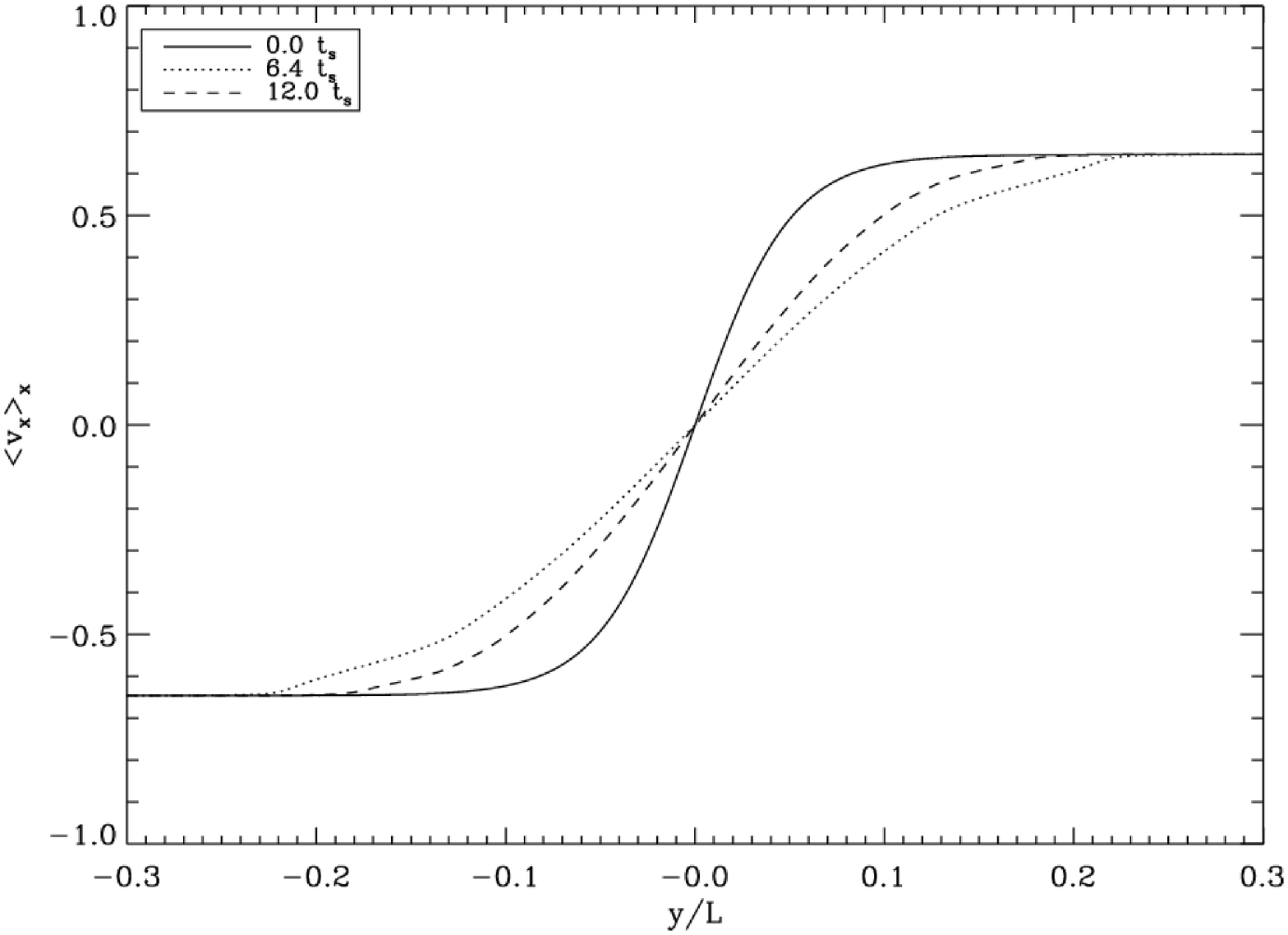} \\
\includegraphics[width=3.5in,angle=90]{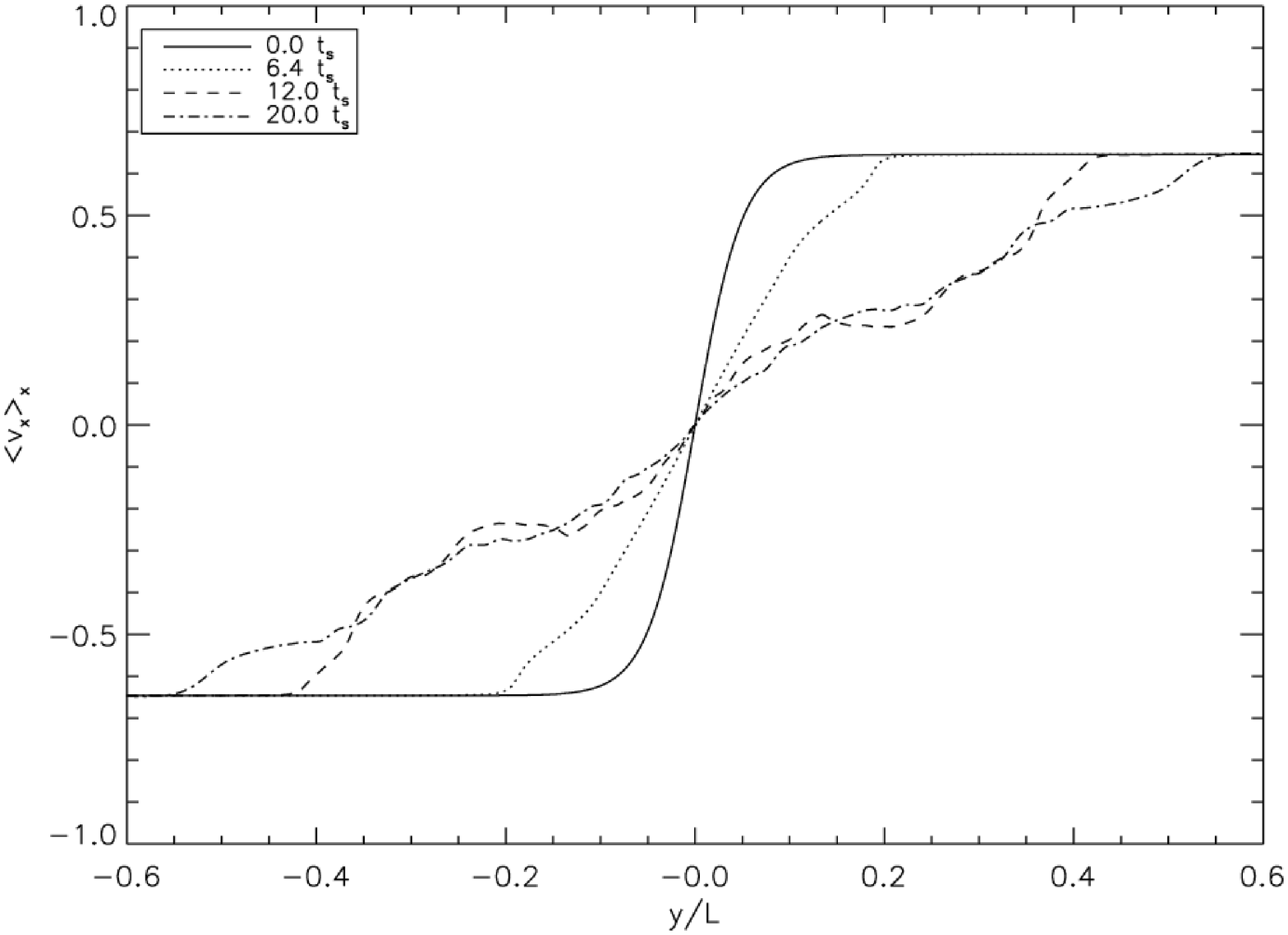}
\caption{The x-averaged x-velocity as a function of y at various times for both the HD (top) and MHD5000 (bottom) models.}
\label{fig:profile}
\end{figure}

As the instability evolves and $KE_{y}$ increases, momentum is transported between the upper ($y>0$) and lower ($y<0$) domains.  As a result, the initial sheared velocity profile (see eq.~\ref{e:profile}) will spread out.  In Fig.~\ref{fig:profile} we plot the x-averaged x velocity as a function of y at different times for both the HD and MHD5000 models.

The evolution of the profile differs between the two models.  In the HD model, the profile reaches a maximum width of $\sim 0.12$ at saturation, $t=6.4t_{s}$ (see dashed line in Fig. \ref{fig:profile}), and then the width oscillates in a non-uniform fashion (see Fig. \ref{fig:profilewidth}), with a minimum of $\sim 0.08$ at ($t=12t_{s}$).  The average width is $\sim 0.1$. If we take the evolved profile as an initial condition, we find that it is stable (presumably it is unstable to longer wavelength modes that are not captured within our periodic domain).

 The profile width in the MHD model, however, continues to grow after saturation, finally reaching its maximum of $~0.37$ at $t\sim 15t_{s}$, when the perturbations have nearly dissipated, and remaining constant thereafter. The spatial roughness of the MHD case, which contrasts with the smooth profile in the HD case, is due to the
prominent small scale
flow structure in the MHD case. Note that only a portion of the domain is plotted in each figure; the shear layer has not reached the top and
bottom boundaries of the domain, and the momentum flux escaping through these boundaries is found to be small.

\begin{figure}
\centering
\includegraphics[width=3.5in,angle=90]{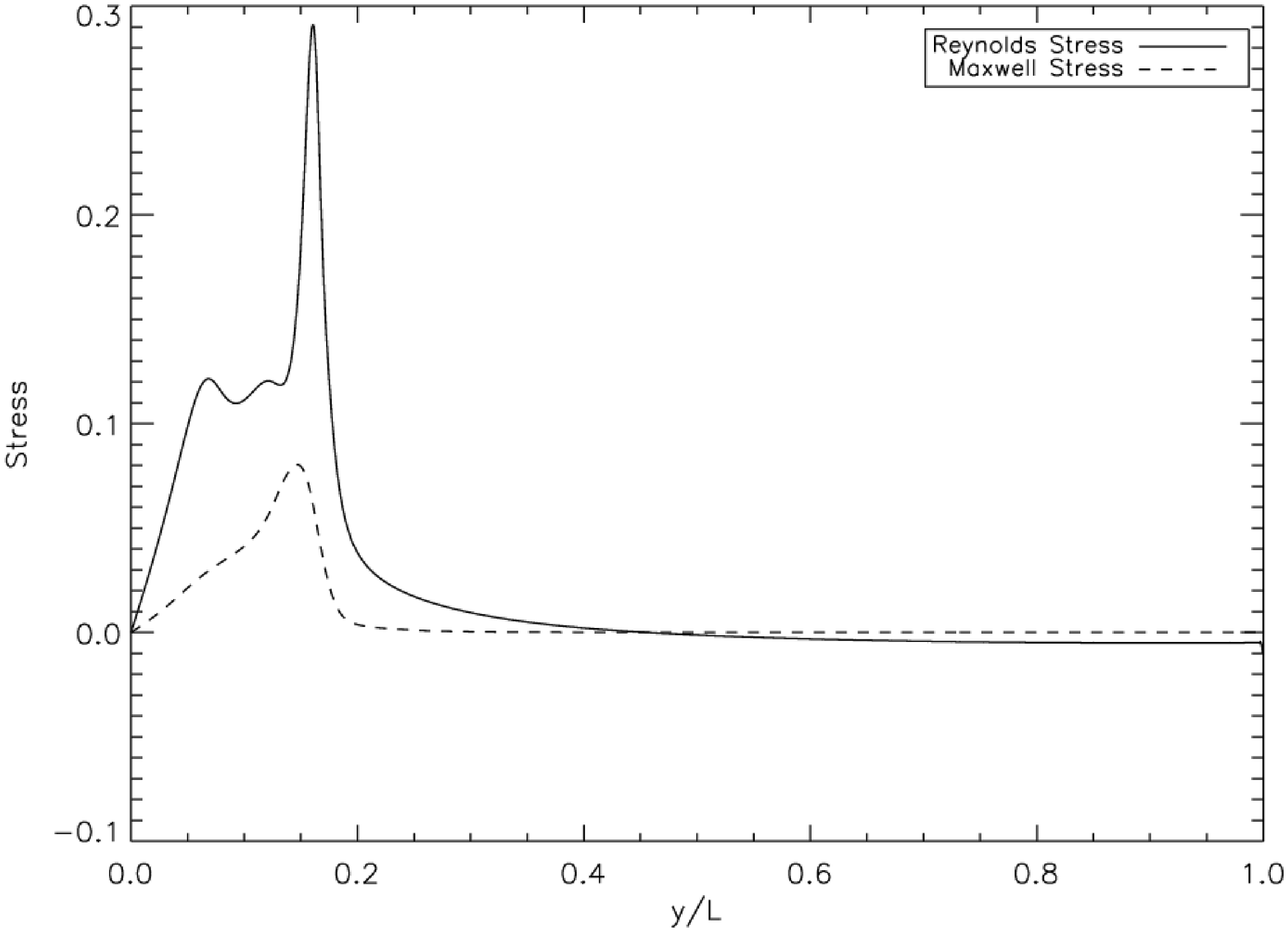} \\
\includegraphics[width=3.5in,angle=90]{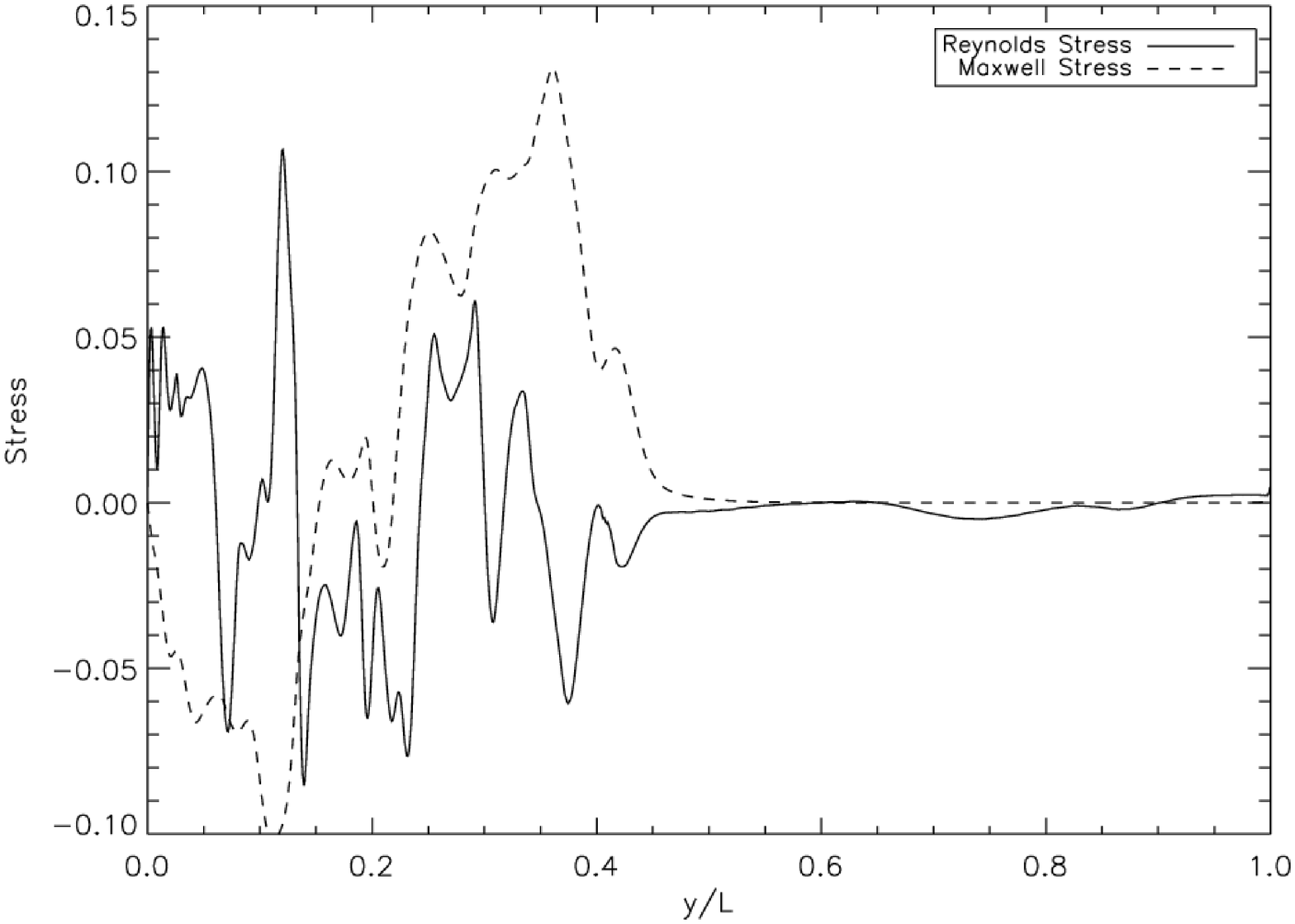}
\caption{The Reynolds and Maxwell stress as a function of y for six times at $R_m=5000$.  The times represented are 6 (top) and 12 (bottom) $t_s$.}
\label{fig:stress}
\end{figure}

In order to understand why the two models differ, we average the x component of the momentum equation (eq.\ref{e:momentum}) over x, getting
\begin{equation}
 \f{\partial \langle\rho v_{x}\rangle}{\partial t} = \f{\partial}{\partial y} (\langle\rho v_x v_y\rangle - \langle B_x B_y\rangle)
\end{equation}
where the first term on the RHS represents Reynolds stress and the second term represents Maxwell stress.  We plot each of these components for the MHD model at various times in Fig. \ref{fig:stress}.  Initially, the Reynolds term dominates the spreading of the layer, reaching a maximum value at saturation ($t=6.4t_{s}$) and then starts to decrease in magnitude, following the HD model.  However, the Maxwell stress term takes over and further broadens the layer until $t\sim 15$ when the broadening ceases. In both cases, the speed at which the layer broadens is subsonic.

\begin{figure}
\centering
\includegraphics[width=3.5in,angle=90]{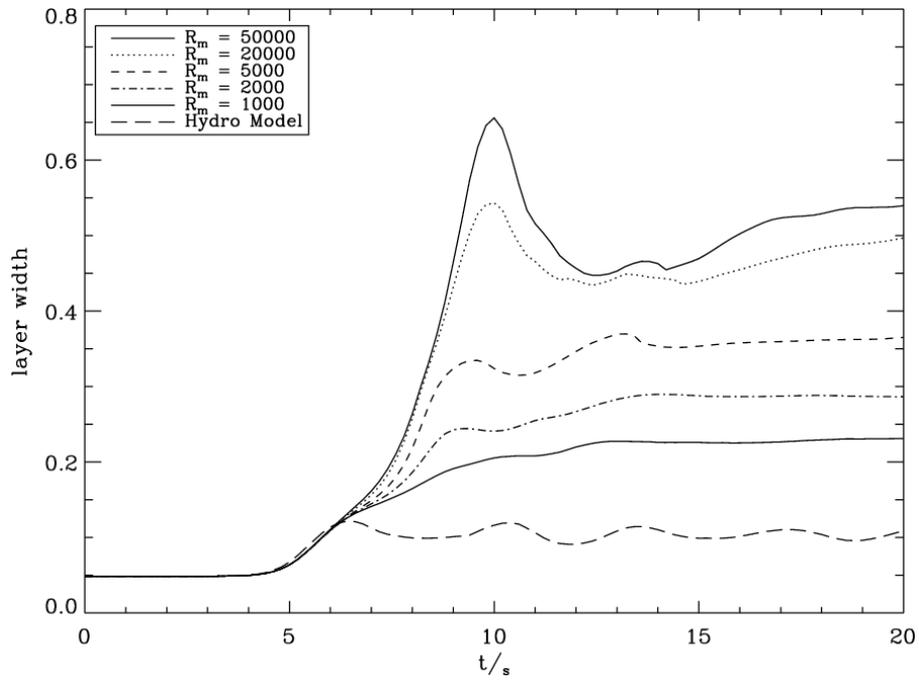}
\caption{The width of the velocity profile as a function of time for magnetic Reynolds number ranging from
 $R_m=1000$ to $R_m=50000$.}
\label{fig:widthall}
\end{figure}

Because in the MHD case the layer is broadened by turbulent stresses, and the amplitude and decay rate of the turbulence depend on $R_m$, we expect the
evolution of layer width to depend on $R_m$ as well. 
This is borne out by Fig. (\ref{fig:widthall}). The
layer width increases with $R_m$, and, like the decay rate, shows no sign of leveling off as $R_m$ increases. Thus, momentum is transported more
efficiently in an MHD model compared to an HD model, and the efficiency increases with increasing $R_m$. Our range of $R_m$ is too small and too
far from ISM values to reliably extrapolate.

\section{SUMMARY AND CONCLUSIONS}\label{s:conclusion}
We have run 2D simulations exploring the non-linear evolution of the Kelvin Helmholtz Instability in both a magnetized and unmagnetized shear flow.  We concentrated on two models:  a hydrodynamic model at sound Mach number, $M_{s}=1$, and a weak-field magnetohydrodynamic model with sound and Alfven Mach numbers, $M_{s}=1$, $M_{a}=10$. We have considered the effect of resistivity by running models with $R_{m}$ ranging from $R_{m}=1000-50000$, at resolutions from $256\times512$ to $2048\times4096$.

Justification for restricting the problem to two dimensions was provided in part by our parameter study of the linear regime: using a linear eigenmode code, we showed that the 2D growthrate is the fastest growing mode for the parameters we considered.  However, the growthrate does not give any information as to the non-linear evolution of the flow, so it is possible that 3D effects would become important in the nonlinear regime.

\begin{figure}
\centering
\includegraphics[width=3in]{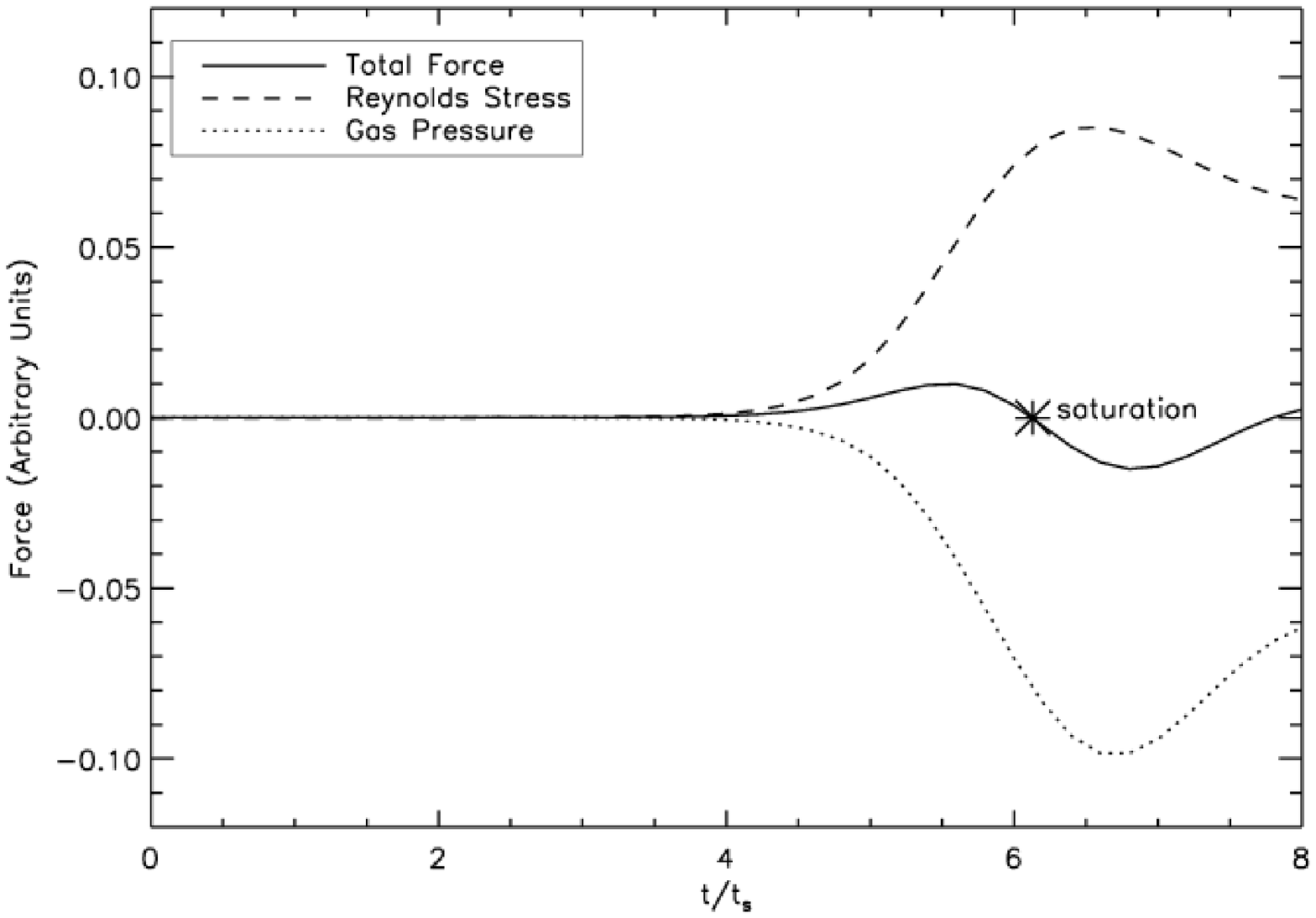} \\
\includegraphics[width=3in]{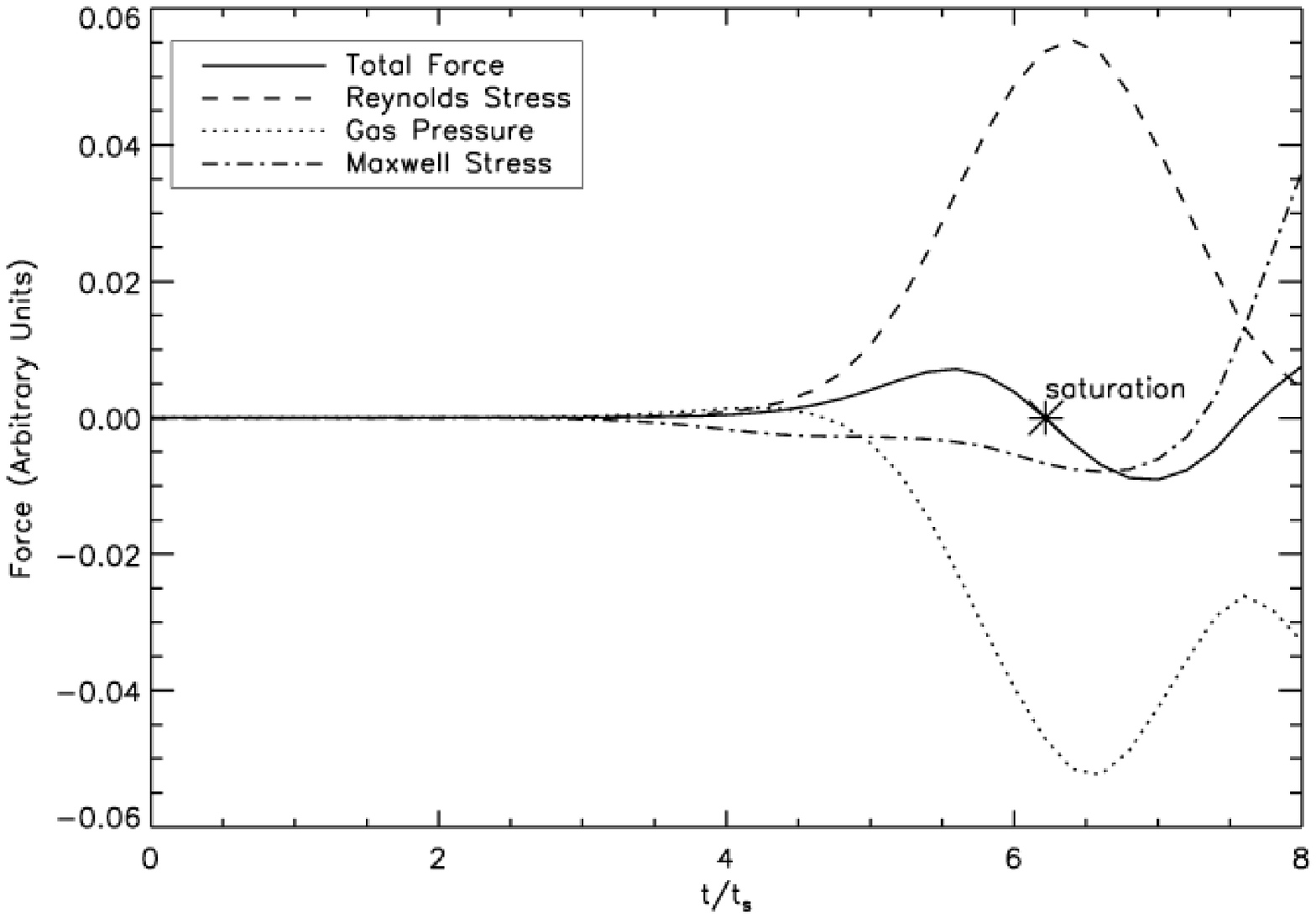}
\caption{The total Reynolds stress, gas pressure, and Maxwell stress as functions of time
for the HD (top) and MHD50000 1024x2048 resolution (bottom) models.  Saturation occurs when all the forces cancel each other out.}
\label{fig:saturation}
\end{figure}

The linear growth phase and the saturation mechanism of a HD flow and a weak field MHD flow show many similarities.  We calculated the growthrates to be different by only $\sim 2.5\%$ between the HD and MHD50000 models.  The mechanism for saturation is an interplay between the Reynolds stress trying to spread the layer and the combined effort of gas pressure and Maxwell stress (in MHD only) trying to prevent the spreading (see Fig.~\ref{fig:saturation}).  We found that the gas pressure dominates over the Maxwell stress in preventing the spreading of the layer.  However, as the initial magnetic field increases, the contribution of Maxwell stress will become more important in saturating the instability, and, for a sufficiently large magnetic field, will suppress it entirely.

\begin{figure}
\centering
\includegraphics[width=2.5in,angle=90]{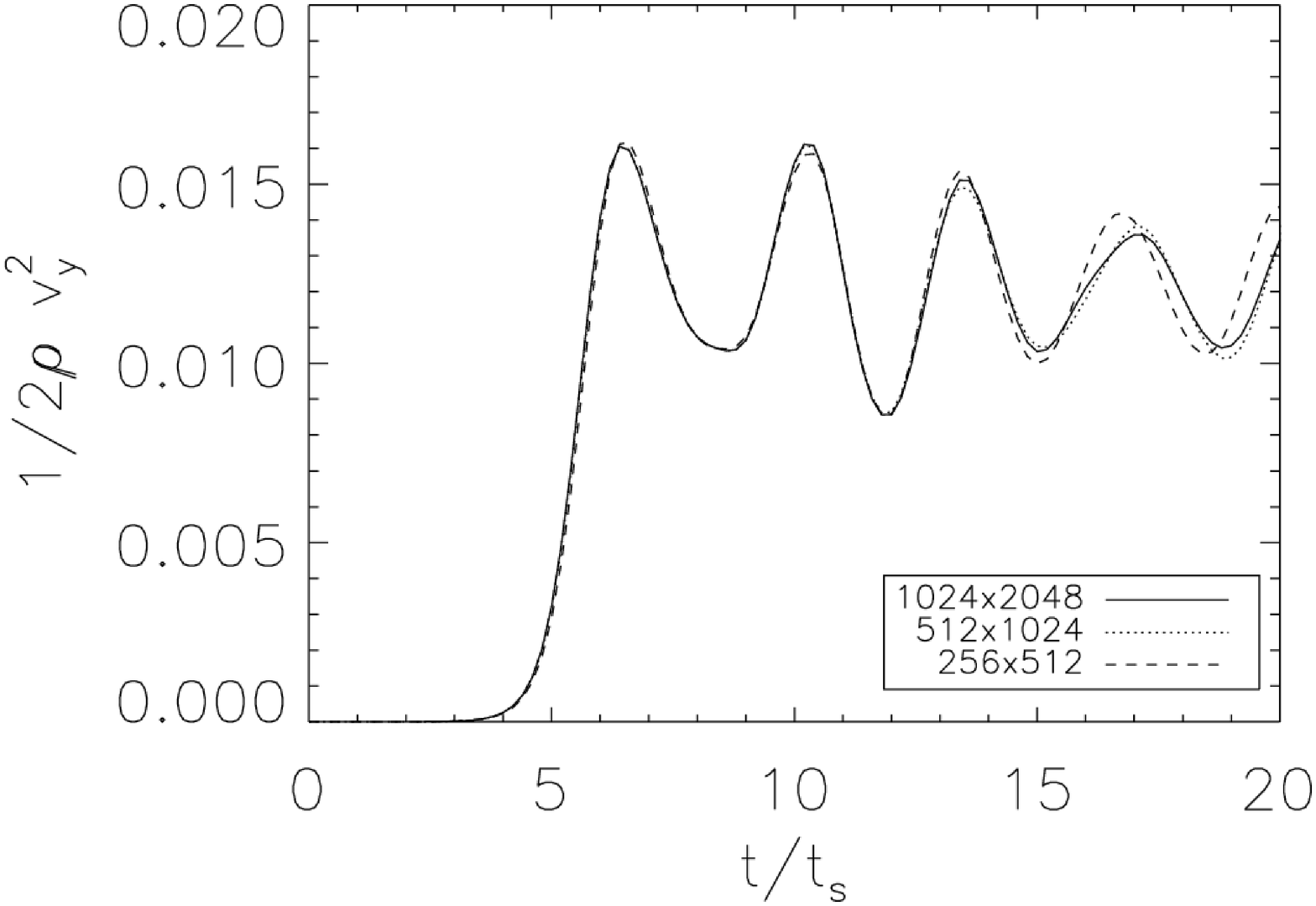} \\
\includegraphics[width=2.5in,angle=90]{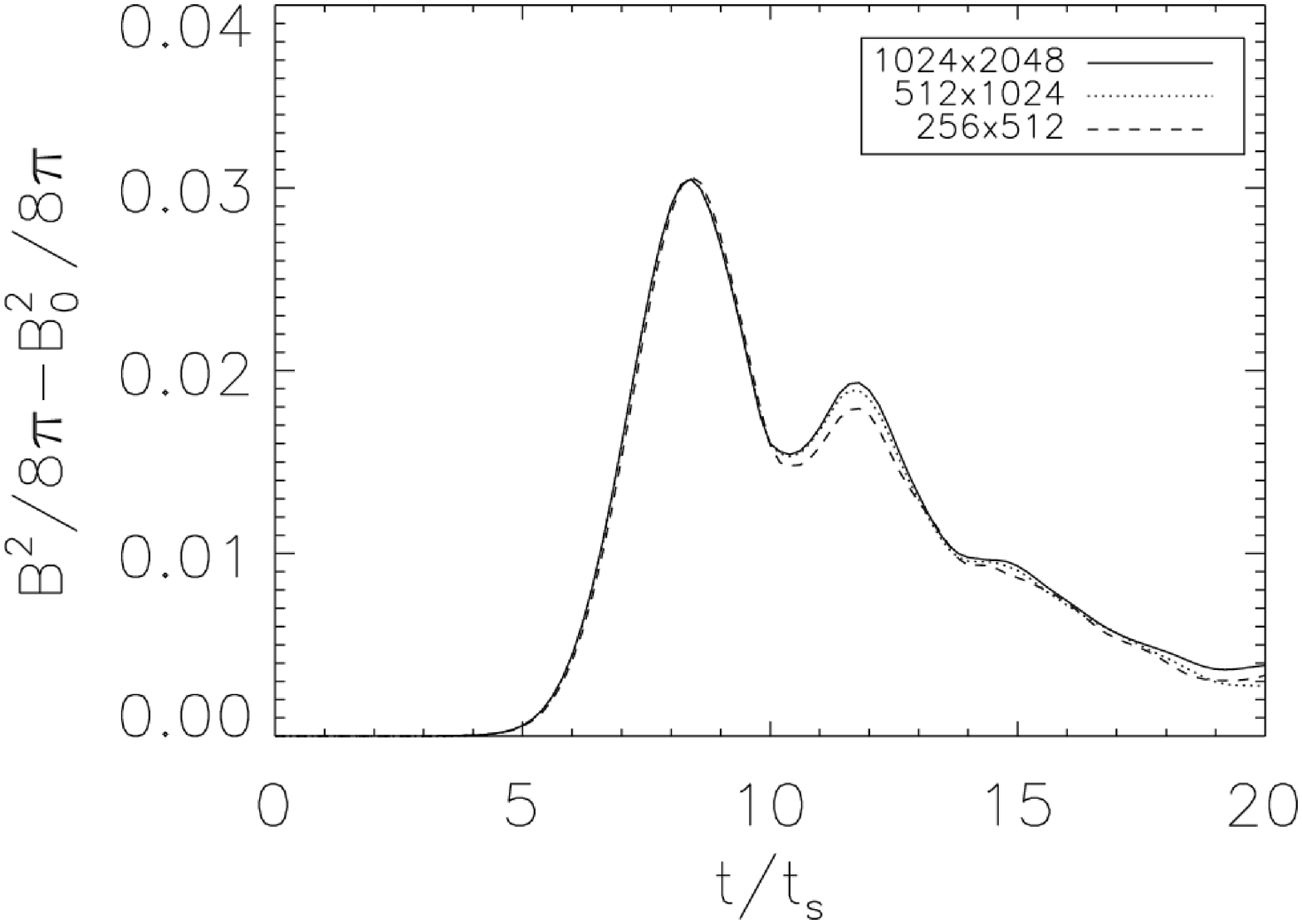} \\
\includegraphics[width=2.5in,angle=90]{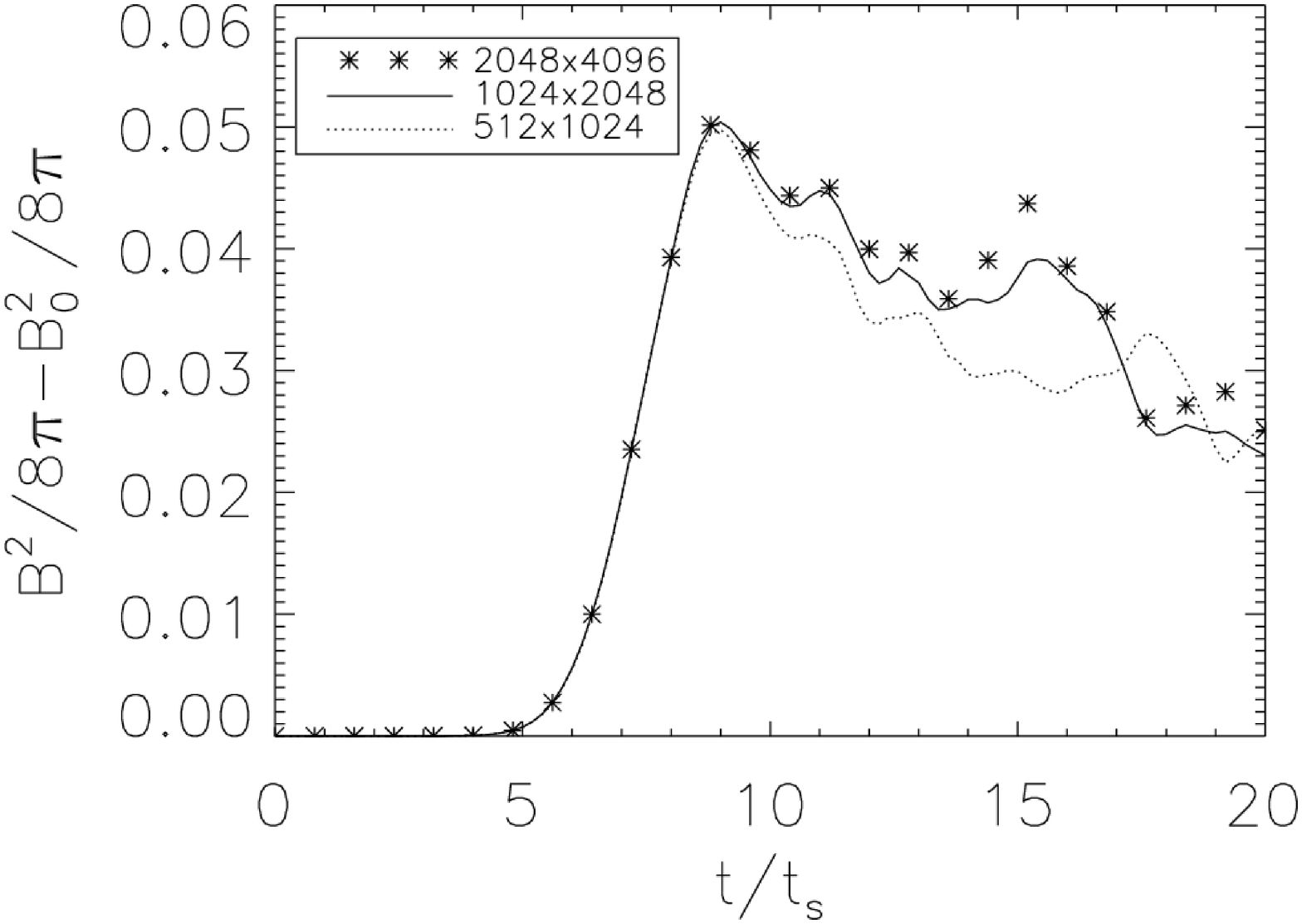} \\
\caption{The total kinetic energy in the y direction as a function of time for the HD model (top) and the perturbed
magnetic energy as  functions of time for model
MHD5000 (middle) and MHD50000 (bottom) are plotted for three resolutions.}
\label{fig:convergence}
\end{figure}

\begin{figure}
\centering
\includegraphics[width=3in]{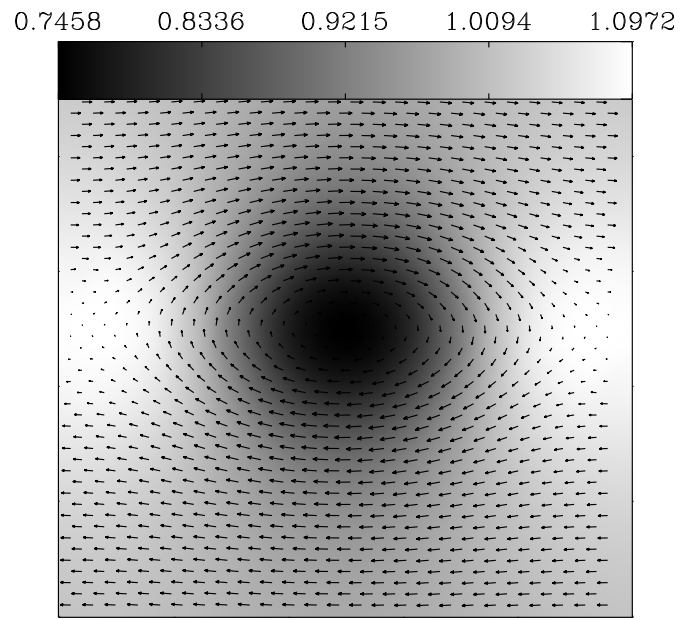}
\caption{Gas density for the high resolution hydrodynamic model after saturation.  The velocity vector field is overplotted.}
\label{fig:hdmorph}
\end{figure}

\begin{figure}
\centering
$\begin{array}{l@{\hspace{1in}}r}
\includegraphics[width=2.5in]{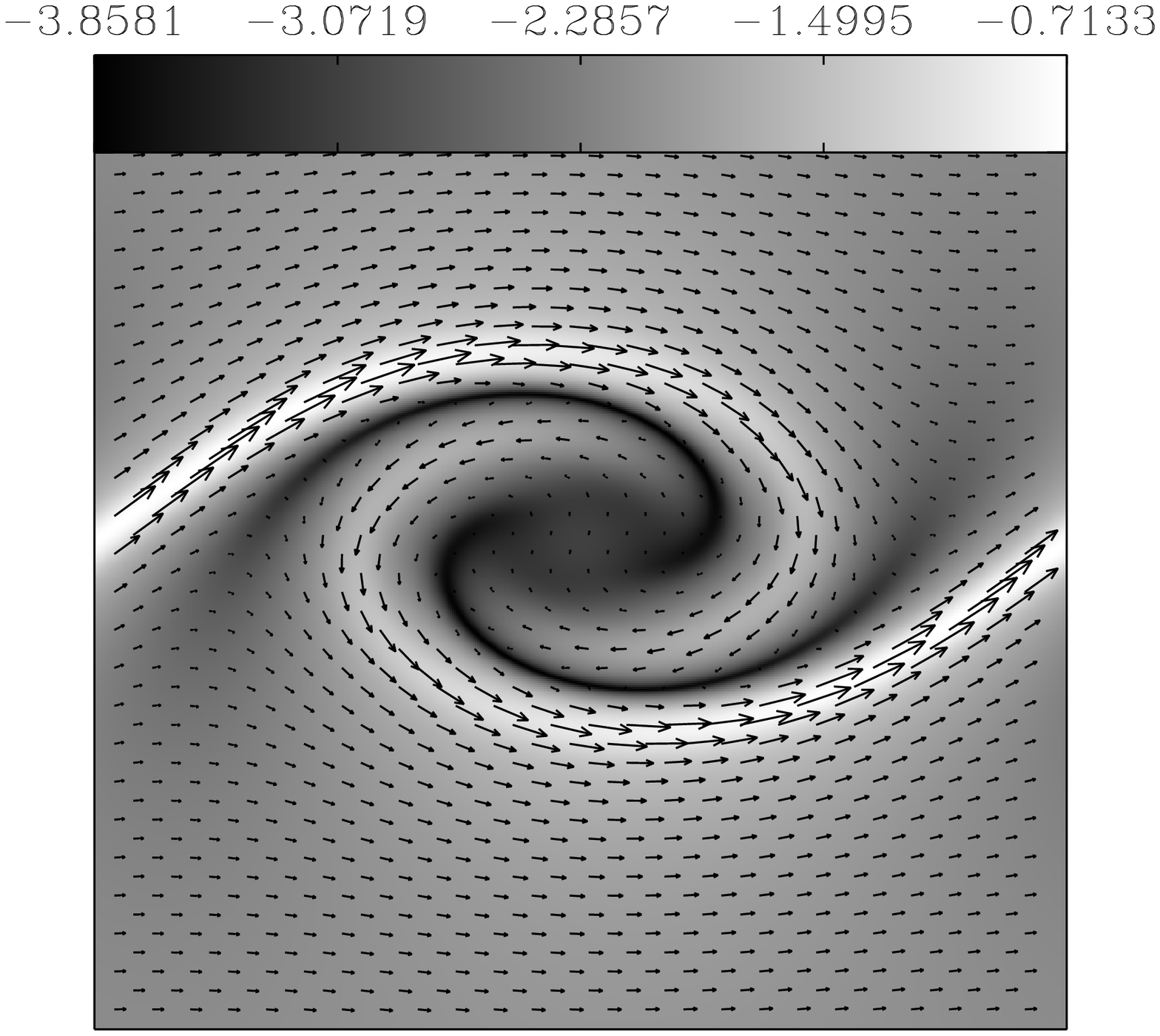} &
\includegraphics[width=2.5in]{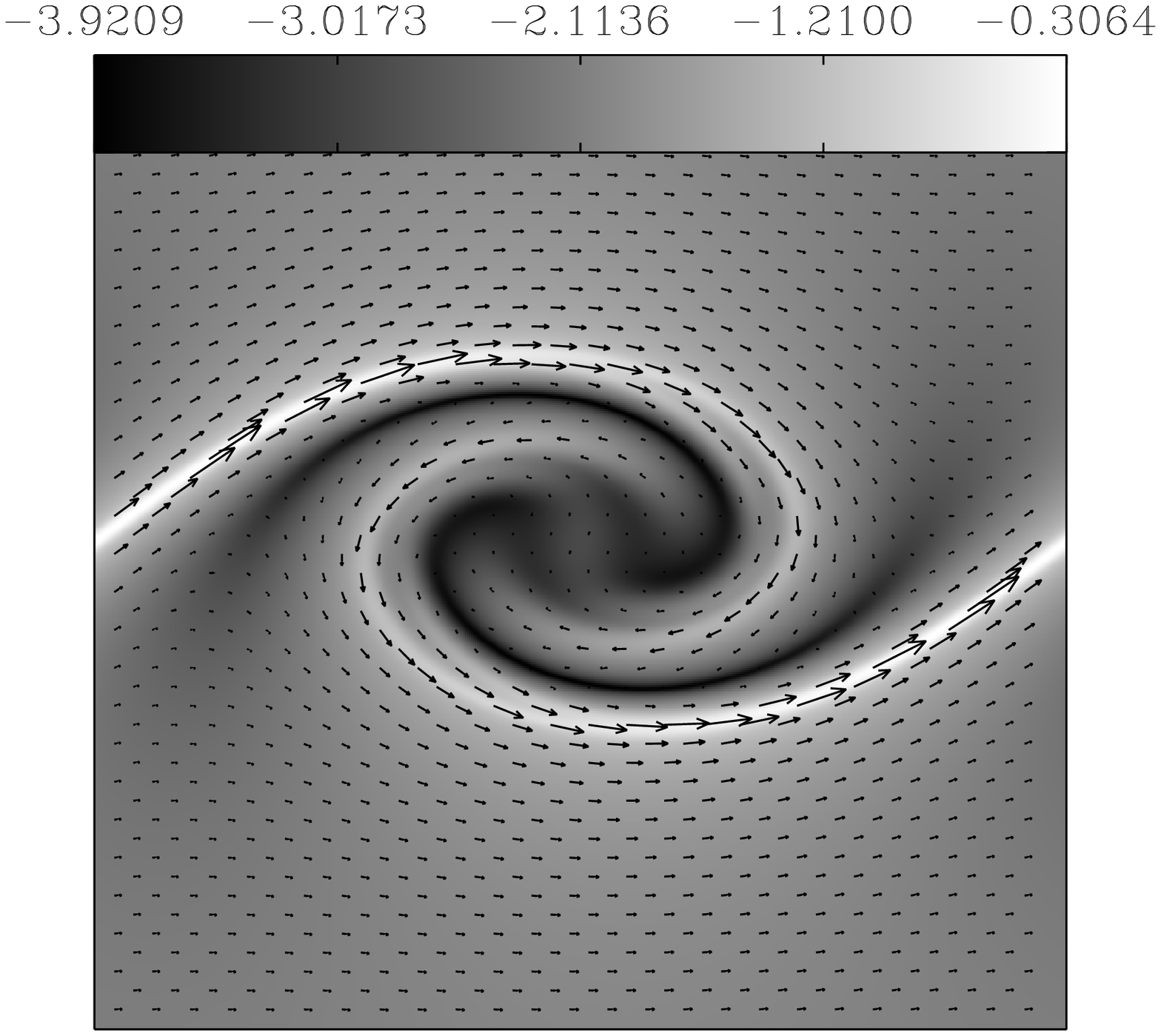} \\
              \\
\includegraphics[width=2.5in]{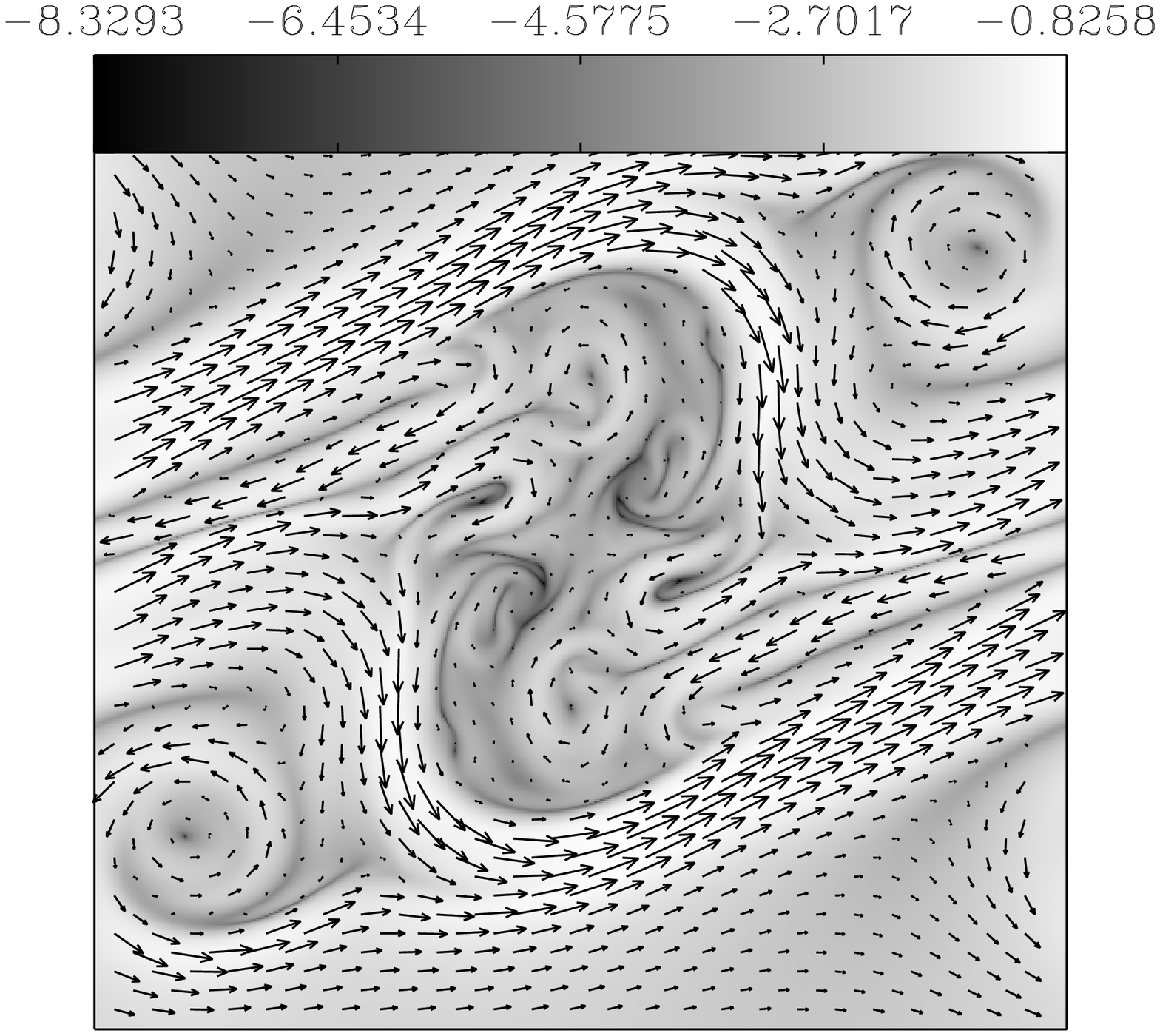} &
\includegraphics[width=2.5in]{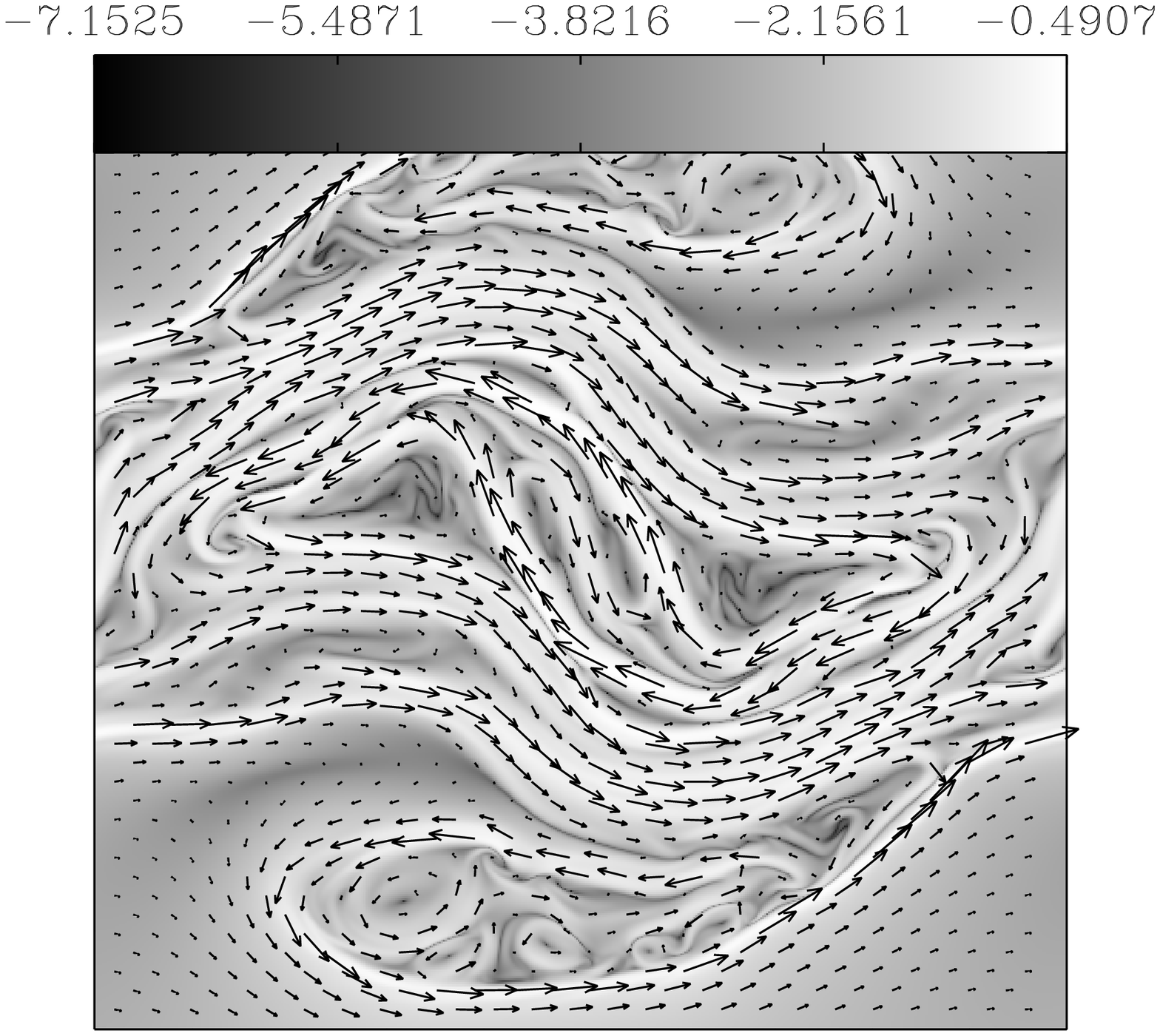} \\
              \\
\includegraphics[width=2.5in]{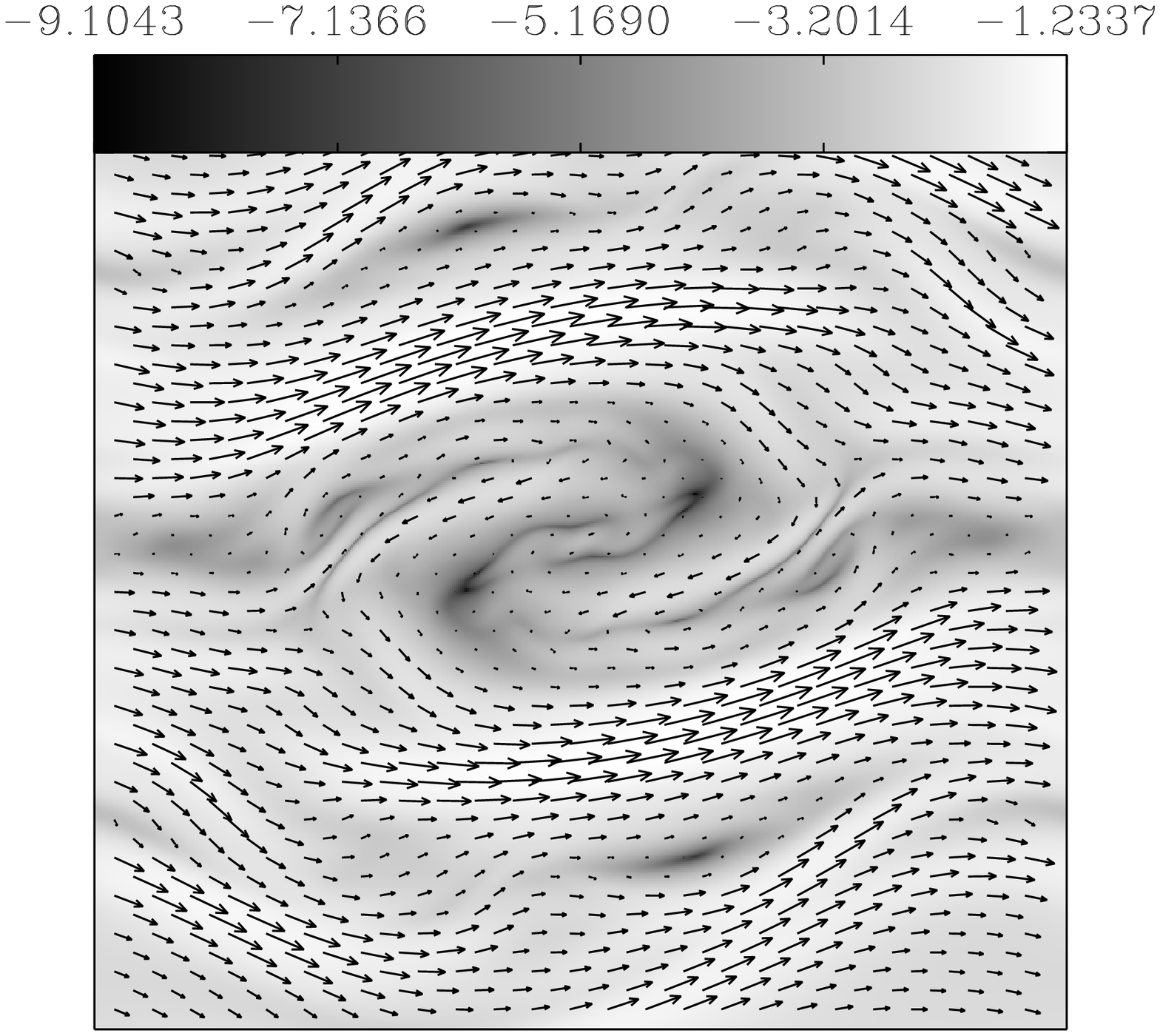} &
\includegraphics[width=2.5in]{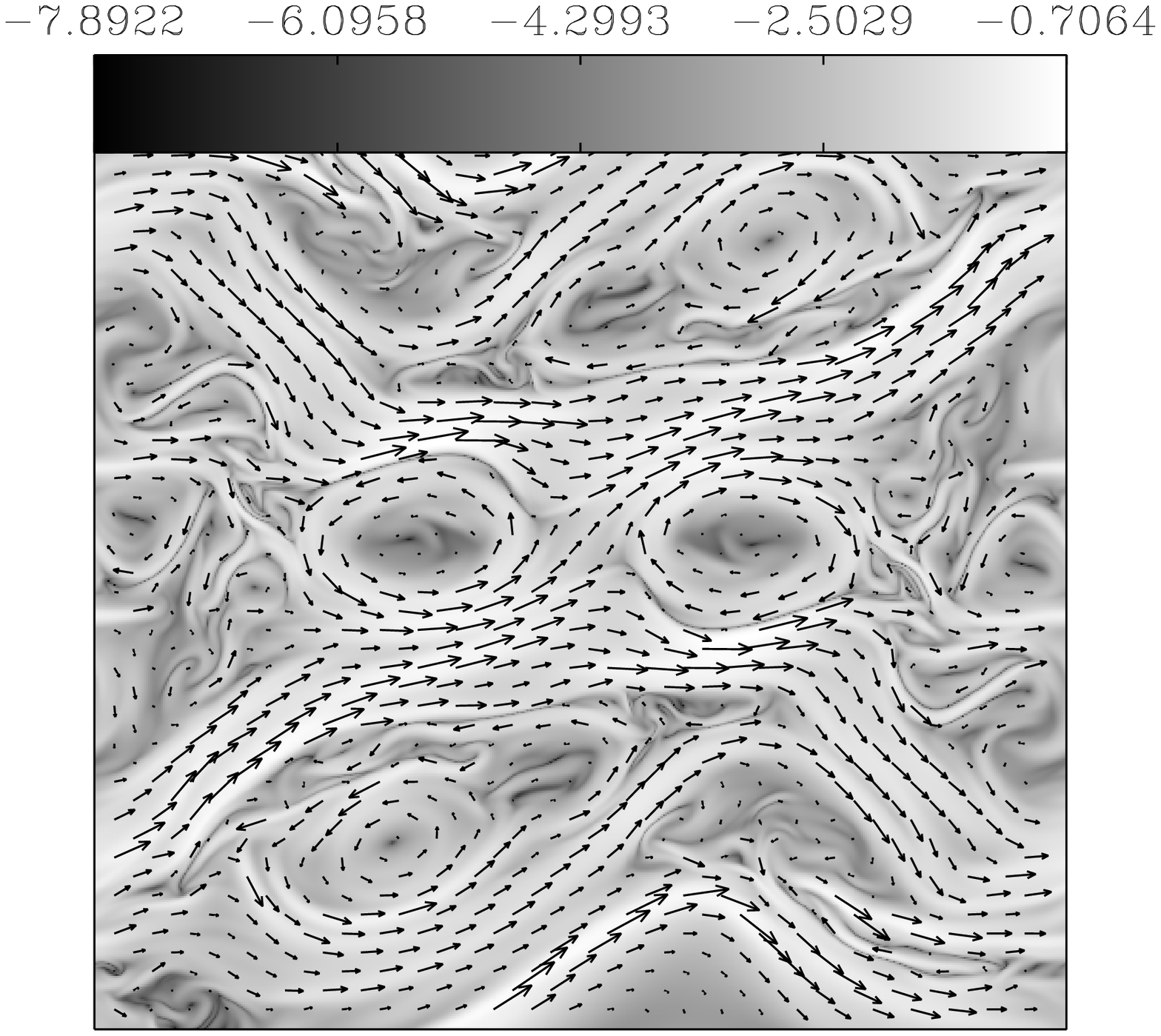}
\end{array}$
\caption{Logarithm of the magnetic energy for the high resolution $R_m$=5000 ({\em left}) and $R_m$=50000 ({\em right}) at three times after saturation.  The magnetic vector field is overplotted.  The times shown are 6.4, 12, and 20 $t_s$.}
\label{fig:mhdme}
\end{figure}

\begin{figure}
\centering
\includegraphics[width=3.5in,angle=90]{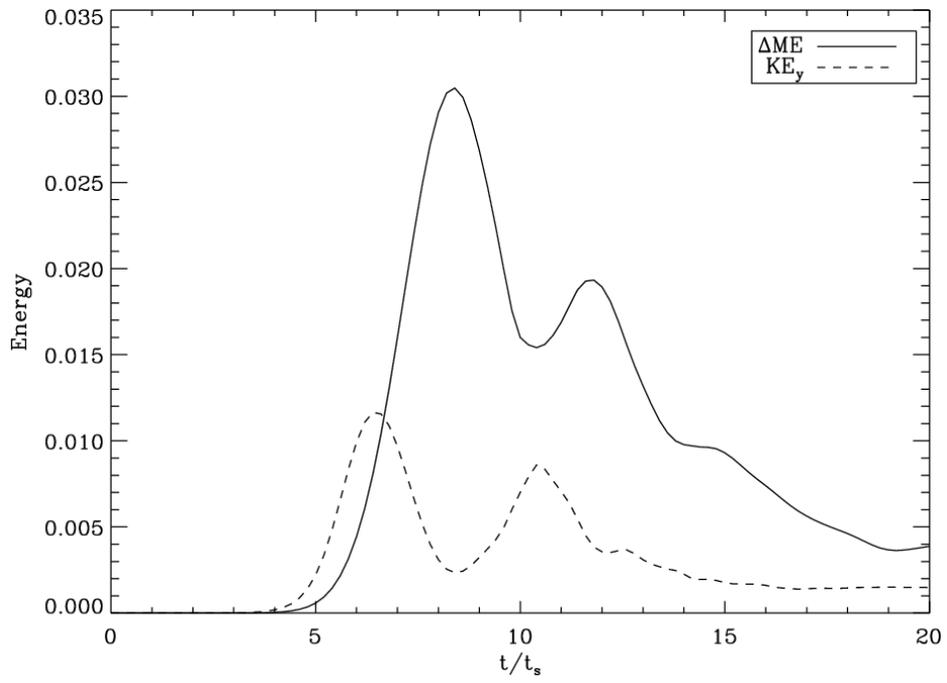}
\caption{ The $y$-component of the kinetic energy, $KE_y$, and the perturbed magnetic energy $\Delta ME$ for $R_m=5000$.}
\label{fig:mhdenergy}
\end{figure}

Despite these similarities in the initial phase of growth, the non-linear evolution between the two models is drastically different (see \S \ref{s:morphology}), as has been known for some time (Frank et al. 1996, Malagoli et al. 1996).  The differences are evident in both the evolution of the kinetic and magnetic energies (top panel of Fig. \ref{fig:convergence} for HD and Fig. \ref{fig:mhdenergy} for MHD) and in the HD density structure (Figs. \ref{fig:hdmorph}) and MHD magnetic structure (Fig. \ref{fig:mhdme}).  The HD flow develops a single large tumbling eddy that remains for the duration of the runs (viscosity will eventually damp it out).  In the MHD flow, the eddy winds up the magnetic field. Magnetic forces spawn a rich array of small scale flow, density, and
magnetic field structures. As the resistivity is lowered, the energy in the fluctuations increases (although it is probably bounded by the
equipartition value) and the decay time of the fluctuations increases. Previous works have suggested that resistive dissipation is the dominant process in the evolution of the flow.  However, if the decay time of the fluctions continue to increase for resistivites lower than considered here, a process with a larger diffusion coefficient may become the dominant process in controlling the evolution of the energy in the flow.  In the ISM, processes with larger diffusion coefficients include both viscosity and ion-neutral friction.  Including these processes is beyond the scope of this work.

The decay rate of the turbulence, and the mode of decay, is not just of academic interest. We showed that the shear layer is broadened by
turbulent Maxwell stresses well beyond what occurs in the hydrodynamic case. The longer the turbulence survives, and the larger its amplitude,
the broader the layer becomes. This means that momentum transport increases with decreasing resistivity.

Although the turbulent decay rate and the momentum transport rate are sensitive to $R_m$ over the range accessible to our study, other properties
of the instability are relatively insensitive. The instability growth rate, times of saturation, and saturated amplitude are relatively
independent of $R_m$, primarily because the magnetic field is too weak to affect these properties. Nevertheless, the dependence on $R_m$ found
for some of the important quantities suggest that extrapolation of these results to real astrophysical flows should be selective and requires some
caution.
\acknowledgements
The authors are happy to acknowledge support from NSF grants AST0507367 and PHY0215581 to the University of Wisconsin and NASA grant NNG06GJ32G to the University of Michigan. 
Portions of this paper were written at the KITP at UC Santa Barbara; we thank them for hospitality and support under NSF grant 0551164.  The ``super high'' resolution models were run at the NCSA (AST060031 to F.H.), all other models were computed on a local PC-cluster, built and maintained by S.~Jansen.

\clearpage

%table 1
\begin{deluxetable}{lcccccc}
\tablecolumns{7}
\tablewidth{0pc}
\tablecaption{Models Run
\label{t:models}}
\tablehead{
\colhead{Model} & \colhead{$M_{s}$} & \colhead{$M_{a}$} & \colhead{$\alpha$} & \colhead{$\beta$} & \colhead{$R_{m}$} & \colhead{Res}
           }
\startdata
HD        & 1 & $\infty$ & 0   & $\infty$ & 0     & l,m,h    \\
MHD1000   & 1 & 10       & 0.1 & 120      & 1000  & l,m      \\
MHD2000   & 1 & 10       & 0.1 & 120      & 2000  & l,m      \\
MHD5000   & 1 & 10       & 0.1 & 120      & 5000  & l,m,h    \\
MHD20000  & 1 & 10       & 0.1 & 120      & 20000 & l,m,h,sh \\
MHD50000  & 1 & 10       & 0.1 & 120      & 50000 & l,m,h,sh \\
\enddata
\end{deluxetable}

\clearpage
%figure 1

%figure 2

%figure 3

%figure 4

%figure 5

%figure 6

%figure 7

%figure 8

%figure 9

%figure 10

%figure 11

%figure 12

%figure 13

%figure 14

\end{document}